\documentclass[twocolumn,preprintnumbers,a4paper,
superscriptaddress,floatfix,twoside,prd]{revtex4}

\usepackage{bm}
\usepackage{epsfig}
\usepackage{graphics}
\usepackage{natbib}

\usepackage{fancyh}
\usepackage[l]{floatflt}
\usepackage{epsfig}
\usepackage{amssymb}
\usepackage{latexsym}
\usepackage{times}
\usepackage{slashed}
\usepackage{upgreek}
\usepackage{amsmath}
\usepackage{amsfonts}
\usepackage{amsbsy}
\usepackage{amscd}
\usepackage{bbm}

\newcommand{\Eref}[1]{Eq.~(\ref{#1})}
\newcommand{\Sref}[1]{Sec.~\ref{#1}}
\newcommand{\Fref}[1]{Fig.~\ref{#1}}
\newcommand{\Tref}[1]{Table~\ref{#1}}
\newcommand{\cref}[1]{Ref.~\cite{#1}}

\newcommand{\what}{\ensuremath{\widehat}}

\newcommand{\astroph}[1]{{\ftn\tt astro-ph/#1}}
\newcommand{\arxiv}[1]{{\ftn\tt  arXiv:#1}}

\newcommand{\bal}{\begin{align}}
\newcommand{\eal}{\end{align}}
\newcommand{\beqs}{\begin{subequations}}
\newcommand{\eeqs}{\end{subequations}}
\newcommand{\eec}{\end{center}}
\newcommand{\bec}{\begin{center}}

\newcommand{\eem}{\end{matrix}}
\newcommand{\bem}{\begin{matrix}}
\newcommand{\eeq}{\end{equation}}
\newcommand{\beq}{\begin{equation}}
\newcommand{\ba}{\begin{array}}
\newcommand{\ea}{\end{array}}
\newcommand{\bea}{\begin{eqnarray}}
\newcommand{\eea}{\end{eqnarray}}
\newcommand{\baq}{\begin{eqnarray}}
\newcommand{\eaq}{\end{eqnarray}}

\newcommand\eqs[2]{Eqs.~(\ref{#1}) and (\ref{#2})}

\newcommand\eqss[3]{Eqs.~(\ref{#1}), (\ref{#2}) and (\ref{#3})}

\newcommand{\ftn}{\footnotesize}

\newcommand{\ssz}{\scriptsize}

\newcommand{\GeV}{{\mbox{\rm GeV}}}

\def\to{\rightarrow}

\def\llgm{\left\lgroup}
\def\rrgm{\right\rgroup}
\def\lf{\left(}
\def\rg{\right)}
\newcommand\vev[1]{\langle {#1} \rangle}
\newcommand\vevi[1]{\langle {#1} \rangle_{\rm I}}

\newcommand{\Vhi}{\ensuremath{ V_{\rm E}}}

\newcommand{\Hhi}{\ensuremath{ H_{\rm E}}}
\newcommand{\wrh}{\ensuremath{w_{\rm rh}}}

\newcommand{\Khi}{\ensuremath{K}}
\newcommand{\Whi}{\ensuremath{W}}
\newcommand{\Vhio}{\ensuremath{ V_{\rm E0}}}
\newcommand{\Ns}{\ensuremath{{N_\star}}}

\newcommand{\mP}{\ensuremath{m_{\rm P}}}

\def\openone{\leavevmode\hbox{\small1\kern-3.8pt\normalsize1}}
\newcommand{\dV}{\ensuremath{\Delta V_{\rm E}}}

\newcommand{\fr}{\ensuremath{f_{\rm p}}}

\newcommand{\msn}{\ensuremath{\what m_{\rm \dph}}}

\newcommand{\ks}{\ensuremath{k_\star}}
\newcommand{\ns}{\ensuremath{n_{\rm s}}}

\newcommand{\nb}{\ensuremath{N_S}}
\newcommand{\as}{\ensuremath{a_{\rm s}}}
\newcommand{\As}{\ensuremath{A_{\rm s}}}

\newcommand{\rce}{\ensuremath{{\mathcal{R}}}}
\newcommand{\Ve}{\ensuremath{{V}}}

\newcommand{\mq}{\ensuremath{\what m_{\rm I}}}

\newcommand{\dphi}{\ensuremath{\what{\delta\phi}}}
\newcommand{\dph}{\ensuremath{\delta\phi}}

\def\aal{{\bar\alpha}}
\def\bbet{{\bar\beta}}
\def\al{{\alpha}}

\def\n{\bar{n}}

\def\th{{\theta}}

\newcommand{\Trh}{\ensuremath{T_{\rm rh}}}
\newcommand{\sg}{\ensuremath{\phi}}

\newcommand{\sgx}{\ensuremath{\phi_\star}}
\newcommand{\sgf}{\ensuremath{\phi_{\rm f}}}
\newcommand{\sgm}{\ensuremath{\phi_{\rm mx}}}
\newcommand{\sgn}{\ensuremath{\phi_{\rm mn}}}

\newcommand{\ld}{\ensuremath{\lambda}}
\newcommand{\lda}{\ensuremath{\lambda_1}}
\newcommand{\ldb}{\ensuremath{\lambda_2}}
\newcommand{\mma}{\ensuremath{M_1}}
\newcommand{\mmb}{\ensuremath{M_2}}

\newcommand{\ma}{\ensuremath{\delta{\small\sf EM}}}
\newcommand{\mb}{{\small\sf EM2} and {\small\sf EM4}}
\newcommand{\mba}{{\small\sf EM2}}
\newcommand{\mbb}{{\small\sf EM4}}
\newcommand{\rs}{\ensuremath{\delta_{21}}}
\newcommand{\rss}{\ensuremath{r_{21}}}
\newcommand{\rrs}{\ensuremath{r_{12}}}
\newcommand{\fss}{\ensuremath{f_{21}}}
\newcommand{\ffs}{\ensuremath{f_{12}}}
\newcommand{\Dex}{\ensuremath{\Delta_{\rm \star}}}

\newcommand{\kas}{\ensuremath{K_1}}
\newcommand{\tkas}{\ensuremath{\widetilde K_1}}
\newcommand{\kbas}{\ensuremath{K_{12}}}
\newcommand{\tkbas}{\ensuremath{\widetilde K_{12}}}

\newcommand{\kb}{\ensuremath{K_{2}}}

\newcommand\mtta[4]{\mbox{
$\llgm\bem #1 &#2 \cr #3& #4\eem\rrgm$}}

\newcommand{\se}{\ensuremath{\widehat \phi}}
\newcommand{\sex}{\ensuremath{\widehat{\phi}_\star}}
\newcommand{\sef}{\ensuremath{\widehat{\phi}_{\rm f}}}
\newcommand{\geu}{\ensuremath{ g}}
\newcommand{\eph}{\ensuremath{\epsilon}}
\newcommand{\ith}{\ensuremath{\eta}}

\def\Ka{K\"{a}hler potential}
\def\Km{K\"{a}hler manifold}
\def\Kaa{K\"{a}hler~}
\def\sub{subplanckian}

\newcommand{\plk}{{\it Planck}}
\newcommand{\tmd}{{TMI}}
\newcommand{\emd}{{EMI}}

\newcommand{\diag}{\ensuremath{{\sf diag}}}
\renewcommand{\max}{\ensuremath{{\sf max}}}

\newcommand{\re}{\ensuremath{{\sf Re}}}

\newcommand{\phc}{\ensuremath{\Phi}}
\newcommand{\phcs}{\ensuremath{\Phi^*}}

\renewcommand{\refname}{{\bf\scshape References}}

\renewcommand{\thesubsection}{{\small\sf\Alph{subsection}}}

\renewenvironment{subequations}{%
\refstepcounter{equation}%
\setcounter{parentequation}{\value{equation}}%
  \setcounter{equation}{0}
  \ignorespaces
}{%
  \setcounter{equation}{\value{parentequation}}%
  \ignorespacesafterend
}

\begin{document}

%\preprint{UT-STPD-2/10}Minimalistic An Hyperbolic Geometry A Suitable Framework for Pole Inflation

\title{\bf\scshape An Alternative Framework for E-Model Inflation in Supergravity}

\author{\scshape Constantinos Pallis\\ {\small\it Laboratory of Physics, Faculty of
Engineering, Aristotle University of Thessaloniki, GR-541 24
Thessaloniki, GREECE} \\ {\ftn\sl  e-mail address: }{\ftn\tt
kpallis@gen.auth.gr}}

%\date{\today}

%of the recent proposed models of inflation

\begin{abstract}

\noindent {\ftn \bf\scshape Abstract:} We present novel
realizations of E-model inflation within Supergravity which are
largely associated with the existence of a pole of order one in
the kinetic term of the (gauge-singlet) inflaton superfield. This
pole arises due to the selected logarithmic \Ka s $\kas$ and
$\tkas$, which parameterize the same hyperbolic manifold with
scalar curvature ${\cal R}_{K}=-2/N$, where $(-N)<0$ is the
coefficient of a logarithmic term. The associated superpotential
$W$ exhibits the same $R$ charge with the inflaton-accompanying
superfield and includes all the allowed renormalizable terms. For
$K=\kas$, inflation can be attained for $N=2$ at the cost of some
tuning regarding the coefficients of the $W$ terms and predicts a
tensor-to-scalar ratio $r$ at the level of $0.001$. The tuning can
be totally eluded for $K=\tkas$, which allows for quadratic- and
quartic-like models with $N$ values increasing with $r$ and
spectral index $\ns$ close or even equal to its present central
observational value.
\\ \\ {\scriptsize {\sf PACs numbers: 98.80.Cq, 04.50.Kd, 12.60.Jv, 04.65.+e}
%Modified theories of gravity
%\hfill {\sl\bfseries Published in} {\sl Phys. Rev. D} {\bf 91},
%123508 (2015)

}
%\pacs{98.80.Cq, 11.30.Qc, 11.30.Er, 11.30.Pb, 12.60.Jv} it does not require fine tuned parameters

\end{abstract}\pagestyle{fancyplain}

\maketitle

\rhead[\fancyplain{}{ \bf \thepage}]{\fancyplain{}{\sl An
Alternative Framework for EMI in SUGRA}} \lhead[\fancyplain{}{\sl
C. Pallis}]{\fancyplain{}{\bf \thepage}} \cfoot{}

\section{Introduction}\label{intro}

It is well-known \cite{terada,pole,new,sor} that the presence of a
pole in the kinetic term of the inflaton gives rise to
inflationary models collectively named $\alpha$-attractors
\cite{alinde, eno5, eno7, class, tkref, tkin, nsreview, linde21,
ellis21}. These models can be classified \cite{tkref, tkin,
nsreview, linde21, ellis21, plin} into \emph{E-Model Inflation}
({\sf\ftn EMI}) \cite{alinde} (or $\alpha$-Starobinsky model
\cite{eno7}) and \emph{T-Model Inflation} ({\sf\ftn \tmd})
\cite{tmodel}, depending on the form of the inflationary
potential, $V_\alpha$, expressed in terms of the canonically
normalized inflaton $\se$. Namely, $V_\alpha$ can be defined as
follows
\beq V_\alpha=\begin{cases}V_{\rm E}\lf1-e^{\sqrt{2/N}\se}\rg& \mbox{for \emd,}\\
V_{\rm T}\lf\tanh{\lf\se/\sqrt{2N}\rg}\rg& \mbox{for \tmd,}
\end{cases}\label{etm}\eeq
where $N>0$ and $V_{\rm E,T}=V_{\rm E,T}(\sg)$ can be polynomial
functions of the initial (non-canonical) inflaton $\sg$.  Given
that the relation between $\sg$ and $\se$ for a kinetic pole of
order {\it two} is $\sg\sim \tanh\se$, the form for $V_{\rm T}$ in
\Eref{etm} can be easily achieved replacing just $\sg$ into a
simple power-law potential $V_{\rm T}=V_{\rm T}(\sg)$
\cite{tmodel}. To process similarly \emd, one would expect, that
the $\sg-\se$ relation would be
\beq \sg=1-e^{-\sqrt{2/N}\se}.\label{se0}\eeq
Up-to-now, to our knowledge, the unity is omitted from the
relation above and so, the attaintment of \emd\ is less
straightforward than the one of \tmd -- see, e.g., \cref{linde21}.
We below investigate the ramifications generated by the inclusion
of unity in \Eref{se0} within \emph{Supergravity} ({\sf\ftn
SUGRA}).

In the latter context, the achievement of $V_\alpha$ from the
F--term SUGRA potential, $V_{\rm F}$, depends on both the
superpotential, $W$, and the \Ka, $K$, governing the singular
kinetic mixing in the inflaton sector. Both $\alpha$-attractor
models are widely connected with the hyperbolic \Kaa\ geometry of
the Poincar\'e disk \cite{sky} or the no-scale SUGRA \cite{eno7,
class, nsreview} -- for alternative geometries see \cref{other}.
These schemes, however, do not produce the $\sg-\se$ relation in
\Eref{se0} and so, a special selection of $W$ is required for the
attaintment of \emd, invoking motivations from the string theory
\cite{eno7, class} or the superconformal formulation of SUGRA
\cite{alinde, tkin, tkref}.

Another difficulty involved in modelling E- and T-model inflation
is relied on the fact that the pole appearing in the inflaton
kinetic term is generically expected to emerge also in $V_{\rm
F}$. The successful implementation of inflation, however, is
smoothen eliminating the pole from $V_{\rm F}$. This aim can be
accomplished \cite{sor} by two alternative
strategies:\vspace*{-0.2cm}

\subparagraph{\sf\sl (a)} Adjusting $W$ and constraining the
curvature of the \Km, so that the pole is removed from $V_{\rm F}$
thanks to cancellations \cite{eno5, eno7, sor} which introduce
some tuning, though.

\subparagraph{\sf\sl (b)} Adopting a more structured $K$ which
yields the usual kinetic terms but remains invisible from $V_{\rm
F}$ \cite{tkref, tkin, sor}. In a such case, any tuning on the $W$
parameters can be eluded.\\

Taking into account the above guidelines, we here establish a
novel realization of \emd, tied to a simple, generic and
renormalizable $W$, consistent with an $R$ symmetry. This becomes
possible thanks to the modified hyperbolic geometry of the
internal space, which displays a pole of order one placed far away
from the origin -- cf. \cref{class, tkref, tkin}. In a such case,
\Eref{se0} is naturally generated -- see \Sref{sugra1} -- and so,
\emd\ can be automatically processed as is done for \tmd. To be
more specific, we justify in the following the super- and \Ka s,
$W=W(z^\al)$ and $K=K(z^\al,z^{*\aal})$ respectively, adopted in
our proposal

Given that EMI is of chaotic type, we follow the systematic
approach of \cref{rube} which bases the implementation of chaotic
inflation within SUGRA on the introduction of two chiral
gauge-singlet superfields, the inflaton $z^1=\phc$, and the
``stabilizer" or Goldstino  $z^2=S$. Placing $S$ at the origin
during EMI, the derivation and the boundedness of the inflationary
potential is greatly facilitated. The coexistence of $S$ and
$\phc$ in $W$ can be further constrained by invoking an $R$
symmetry under which $S$ and $\Whi$ are equally charged whereas
$\phc$ is uncharged. Contrary to others options
\cite{linde21,ellis21}, where a monomial $W$ is selected, we here
adopt the most general renormalizable $W$, i.e.,
\beq W= S\lf \lda \phc+\ldb\phc^2-M^2\rg \label{whi} \eeq
where $\lda, \ldb$ and $M$ are free parameters. As we see below,
thanks to the general form of $W$ the resulting potential
interpolates between the quadratic and the quartic one causing a
variation to the resulting inflationary observables.

As regards $K$, this includes two terms:\vspace*{-0.2cm}

\subparagraph{\sf \sl (a)} $K_2$ which parameterizes the compact
manifold $SU(2)/U(1)$ and assures the stability of $S$ \emph{with
respect to} ({\sf\ftn w.r.t}) its perturbations during EMI
\cite{su11}
\beq K_{2}=N_S\ln\left(1+|S|^2/N_S\right)
~~\mbox{with}~~0<N_S<6.\label{K2}\eeq\vspace*{-0.6cm}
\subparagraph{\sl (b)} One of the following $K$'s, $\kas$ or
$\tkas$, which yields the desired pole in the kinetic term of
$\phc$ and share the same geometry. Namely,
\beqs \begin{align} \kas&=-N\ln\left(1-(\phc+\phc^*)/2\right),\label{Ks}\\
\tkas&=-N\ln\frac{(1-\phc/2-\phc^*/2)}{(1-\phc)^{1/2}(1-\phc^*)^{1/2}},\label{tkas}\end{align}\eeqs
with $\re\phc<1$ and $N>0$. The placement of the singularity at
unity is crucial in order to obtain the desired $\sg-\se$ relation
in \Eref{se0}. Moreover, these $K$'s admit a well-behaved
expansion for low $\phc$ values. On the other hand, any connection
with no-scale SUGRA \cite{eno7, nsreview} is lost.\\

According the aforementioned discussion, related to the
elimination the pole from $V_{\rm F}$, we can define two versions
of EMI based on the $K$'s above:\vspace*{-0.2cm}

\subparagraph{\sl (a)} $\delta$ \emph{E Model} (\ma) in which we
constrain $N=2$ in \Eref{Ks} and tune
\beq \rss=-\ldb/\lda\simeq 1+\rs\,~~\mbox{and}~~M\ll1
\label{rs}\eeq
in \Eref{whi} such that the denominator including the pole in
$\Vhi$ is (almost) cancelled out. Therefore, \ma\ is relied on
\beq \kbas=\kb+\kas,\label{kbas} \eeq and yields results similar
to the models in \cref{eno5, nsreview} which identify the inflaton
with matter-like superfield.

\subparagraph{\sl (b)} \emph{E Model 2} (\mba) and \emph{E Model
4} (\mbb) which employ \beq \tkbas=\kb+\tkas \label{tkbas}\eeq
with free parameters $N$, $\lda$, $\ldb$ and $M$ and their
discrimination depends on which of the two $\sg$-dependent terms
in \Eref{whi} dominates. Obviously, for $M\simeq0$ and $\ldb=0$ or
$\lda=0$, \mba\ or \mbb\ coincides with the quadratic or quartic
EMI respectively \cite{alinde, plin}. More specifically, \mba\
with $N=3$ yields the potential of the Starobinsky model
\cite{R2}. \\

%As we find, the present models are clearly discriminated from the
%the ones with as regards its geometric structure and its
%observational signatures.

The rest of the paper is organized as follows: In Sec.~\ref{geo}
we describe the geometric structure of $\kas$ and $\tkas$. Then,
in Sec.~\ref{sugra} we describe our inflationary setting and in
Sec.~\ref{res} our models are confronted with observations. Our
conclusions are summarized in Sec.~\ref{con}. Throughout, the
complex scalar components of the various superfields are denoted
by the same superfield symbol, charge conjugation is denoted by a
star ($^*$) -- e.g., $|z|^2=zz^*$ -- the symbol $,z$ as subscript
denotes derivation \emph{with respect to} ({\ftn\sf w.r.t}) $z$,
and we use units where the reduced Planck scale $\mP = 2.44\cdot
10^{18}~\GeV$ is equal to unity.

\section{Geometry of the \Kaa\ Potentials}\label{geo}

At a first glance, both $\kas$ and $\tkas$ are invariant under the
transformation $\Phi \to\ \Phi^*$, whereas $\kas$ is in addition
invariant under the non-holomorphic replacement
\beq \Phi \to\ \Phi+it~~\mbox{with}~~t\in
\mathbb{R}.\label{shift2}\eeq
However, to investigate deeper the symmetries of the $\phc-\phcs$
space and compare it with the geometry of similar models in
\cref{tkref}, we determine its riemannian metric
\beqs\beq \label{ds} ds_K^2=K_{\phc\phc^*} d\phc
d\phc^{*}=\frac{N}{4} \frac{d\phc
d\phc^*}{\lf1-(\phc+\phc^*)/2\rg^2}\eeq and the scalar curvature,
\beq {\cal
R}_{K}=-K^{\phc\phc^*}\partial_\phc\partial_{\phc^*}\ln\lf
K_{\phc\phc^*}\rg=-\frac{2}{N}.\label{curv}\eeq\eeqs
As in the cases of \cref{tkref, eno7} the internal space is also
hyperbolic. On the other hand, $ds_{K}^2$ in \Eref{ds} remains
invariant under the transformation
\beqs\beq \frac{\phc}{2}\to \frac{a \phc/2+b}{c \phc/2+d},
\label{trn}\eeq
with $a,b,c,d\in \mathbb{C}$ and $ab-cd\neq0$, provided that
\beq b=0,~ c=2a,~ d=-a~~\mbox{and}~~|a|^2=1.\label{abcd}\eeq\eeqs
These restrictions signal a different geometry from that of
\cref{tkref, eno7}. Indeed, the matrix ${\bold M}$ representing
the transformation in \eqs{trn}{abcd} has the form
\beq {\bold M}=a\mtta{1}{0}{2}{-1}~~\mbox{with}~~|\det{\bold
M}|^2=1 \label{mt}\eeq
and can be considered as an element of the set of the conjugate
anti-symplectic matrices satisfying the relation
\beq \label{mts} {\bold M}^\dag {\bold E} {\bold M}=-{\bold
E}\>\>\>\mbox{with}\>\>\>{\bold E}=\mtta{0}{1}{-1}{0}\eeq
a nonsingular antisymmetric metric-like matrix. The set of the
matrices ${\bold M}$ does not have a group structure -- e.g., the
identity does not belong to it. However, any matrix ${\bold M}$
can be expressed as the product of a conjugate symplectic matrix
${\bold S}$ times a fixed conjugate anti-symplectic one, such as
the Pauli matrix ${\bold \sigma_3}$. I.e.,
\beq {\bold M}={\bold  S} {\boldmath
\sigma}_3\>\>\>\mbox{with}\>\>\>{\bold \sigma}_3=\diag\lf1,-1\rg.
\label{ms}\eeq
It can be shown \cite{haber} that ${\bold  S}\in U(1,1)$ which is
a Lie subgroup of the M\"obius group, i.e., the projective general
linear transformation group $PGL(2,\mathbb{C})$.

%The Iwasawa-like decomposition of ${\bold M}$ is
%
%\beq {\bold M}= a K \cdot A \cdot N\eeq with
%
%Where the matrices $K, A$ and $N$ parameterize the compact,
%abelian and nilpotent transformations of the M\"obius group. We
%find \bal K=\mtta{1}{-2}{2}{1}, A=\mtta{1}{0}{0}{-\frac{1}{5}},
%N=\mtta{1}{-\frac{2}{5}}{0}{1}.\end{align}
%
Under the transformation in \eqs{trn}{abcd}, $\kas$ in \Eref{Ks}
is transformed as
\beq \label{Kst} \kas\to
\kas+\Lambda+\Lambda^*~~\mbox{with}~~\Lambda=N\ln
\lf1-\phc\rg,\eeq
whereas $\tkas$ in \Eref{tkas} remains precisely invariant, as can
be proven, if we take into account that
\beq \tkas=\kas-\Lambda/2-\Lambda^*/2.\label{tkaskas}\eeq
Note that the corresponding model in \cref{tkref} does not enjoy a
similar invariance under the $SU(1,1)/U(1)$ transformations. To
compare further our framework with that, we introduce the
``canonically normalized'' superfield $X$ (or the Killing--adapted
coordinates) via the relation
\beq \phc=1-e^{-\sqrt{2/N}X}.\label{Xdef}\eeq
Inserting it in \Eref{tkas}, $\tkas$ can be brought into the form
\beq \tkas=-N\ln\cosh\frac{X-X^*}{\sqrt{2N}},\label{Xshift}\eeq
which coincides with the corresponding expression in
$SU(1,1)/U(1)$ geometry \cite{tkref} despite the fact that
\Eref{Xdef} is different. From this result, we conclude that
$\tkas$ is invariant under the shift symmetry
\beq X \to\ X+A ~~\mbox{with}~~A\in\mathbb{R}\label{shift3}\eeq
which is different from that enjoyed by $\kas$ in \Eref{shift2}.
The last equation demonstrates the independence of $\tkas$ from
the canonically normalized inflaton which can be identified as the
real part of $X$.

\section{Inflationary Set-up}\label{sugra}

The (Einstein frame) action within SUGRA for the superfields
$z^\al$'s -- with $\Phi$ ($\al=1$) and $S$ ($\al=2)$ -- can be
written as
\beqs \beq\label{Saction1} {\sf S}=\int d^4x \sqrt{-{
\mathfrak{g}}}\lf-\frac{1}{2}\rce +K_{\al\bbet}
\geu^{\mu\nu}\partial_\mu z^\al \partial_\nu z^{*\bbet}-V_{\rm
F}\rg, \eeq
where $\rce$ is the space-time Ricci scalar curvature,
$\mathfrak{g}$ is the determinant of the background
Friedmann-Robertson-Walker metric, $g^{\mu\nu}$ with signature
$(+,-,-,-)$ and summation is taken over the scalar fields $z^\al$.
Also, \beq K_{\al\bbet}=K_{,z^\al z^{*\bbet}}
~~\mbox{and}~~K^{\al\bbet}K_{\bbet\gamma}=\delta^\al_{\gamma}.\eeq
The F--term SUGRA potential $V_{\rm F}$ is given by
\beq V_{\rm F}=e^{\Khi}\left(K^{\al\bbet}(D_\al W) (D^*_\bbet
W^*)-3{\vert W\vert^2}\right),\label{Vsugra} \eeq \eeqs
where $D_\al W=W_{,z^\al} +K_{,z^\al}W$. In our analysis we use
$W$ in \Eref{whi} and the $K$'s defined in \eqs{Ks}{tkas}. We
below, in \Sref{sugra1}, determine the canonically normalized
fields involved in our setting, derive the inflationary potential
in \Sref{sugra2}, find the SUSY vacuum in \Sref{sugra3} and check
the stability of the inflationary path in \Sref{sugra4}.

\subsection{\small\sf  Canonically Normalized Fields}\label{sugra1}

The inflationary track is determined by the constraints
\beq \label{inftr}
\vevi{S}=\vevi{\Phi-\Phi^*}=0,~\mbox{or}~~\vevi{s}=\vevi{\bar
s}=\vevi{\th}=0\eeq
if we express $\Phi$ and $S$ according to the parametrization
\beq \Phi=\:{\phi\,e^{i \th}}\>\>\>\mbox{and}\>\>\>S=\:(s +i\bar
s)/\sqrt{2}.\label{cannor} \eeq
Here the symbol ``$\vevi{Q}$" denotes the value of a quantity $Q$
during EMI. The canonically (hatted) normalized fields are defined
as follows
\beq \vevi{K_{\al\bbet}}\dot z^\al \dot z^{*\bbet} \simeq
\frac12\lf\dot{\widehat \sg}^2+\dot{\widehat \th}^2+\dot{\what
s}^2+{\dot{}~\widehat{\bar s}}^{~2}\rg, \label{kzzn}\eeq
where the dot denotes derivation w.r.t the cosmic time, $t$.
Taking into account \eqs{Ks}{tkas}, we obtain the diagonal \Kaa\
metric \beq \label{Kab} \lf K_{\al\bbet}\rg=\diag\lf
K_{1\phc\phc^{*}},K_{2SS^*}\rg,\eeq where its elements along the
inflationary path in \Eref{inftr} read \beq\label{Kab1}
\vevi{K_{1\phc\phc^*}}=N/4\fr^2~~\mbox{with}~~\fr=1-\sg~~\mbox{and}~~\vevi{K_{2SS^*}}=1.\eeq
From the first of the relations above we infer that the kinetic
term of $\sg$ exhibits a pole of order one at $\sg=1$. The $\phi$
dependent part of the left-hand side of \Eref{kzzn} is written as
\beq \label{kzz} \vevi{K_{\phc\phc^{*}}}|\dot
\phc|^2=\vevi{K_{\phc\phc^{*}}}\lf\dot
\sg^2+\sg^2\dot\theta^2\rg\,. \eeq
Comparing the last expression with \Eref{kzzn}, in view of
\Eref{Kab1}, we deduce
\beq \label{Je} {d\se}/{d\sg}=J=\sqrt{N/2}\fr^{-1}~~\mbox{and}~~
\widehat{\theta}=J\sg\theta.\eeq
Integrating \Eref{Je} we can identify $\se$ in terms of $\sg$, as
follows
\beq
\se=-\sqrt{N/2}\ln(1-\sg)\>\>\>\Rightarrow\>\>\>\sg=1-e^{-\sqrt{2/N}\se}
,\label{se}\eeq
which is the advertised relation in \Eref{se0}. We remark that
$\se$ gets increased from $0$ to $6.5\sqrt{N}$ for
$0\leq\sg\lesssim0.9999$, i.e., $\se$ can be much larger than
$\sg$ facilitating, thereby, the attainment of EMI with \sub\
$\sg$'s. This output is welcome, since SUGRA can be traditionally
viewed as an effective theory valid for superfield values below
$\mP$.

\subsection{\small\sf  Inflationary Potential}\label{sugra2}

The only surviving term in \Eref{Vsugra} along the trough in
\Eref{inftr} is
\beq \label{Vhi0}\Vhi=\vevi{V_{\rm F}}=\vevi{e^{K}K^{SS^*}
|W_{,S}|^2},\eeq
where \eqs{Ks}{tkas} yield
\beq \label{eK} \vevi{e^{K}}=\begin{cases}
\fr^{-N}&\mbox{for}\>\>\>K=\kbas,\\
1& \mbox{for}\>\>\>K=\tkbas,
\end{cases}\eeq
i.e., the pole in $\fr$ is presumably present in $\Vhi$ of \ma,
but it disappears in $\Vhi$ of \mb, as anticipated by the model's
description in \Sref{intro}. Substituting \eqs{whi}{eK} into
\Eref{Vhi0}, this takes its master form \beq\Vhi=\begin{cases}
%\begin{array}{rl}
%
{\lf \lda \sg+\ldb\sg^2-M^2\rg^2}/{\fr^{N}}&\mbox{for \ma},\\
{\lf \lda \sg+\ldb\sg^2-M^2\rg^2}&\mbox{for
\mb}\,.\end{cases}\label{Vhi}\eeq
The formula above can be brought in the more compact form
\beq\Vhi=\ld^2 \begin{cases}
%\begin{array}{rl}
{\lf  \sg-\rss\sg^2-\mma^2\rg^2}/{\fr^{N}}&\mbox{for \ma},\\
{\lf \sg-\rss\sg^2-\mma^2\rg^2}&\mbox{for \mba},\\
{\lf \sg^2-\rrs\sg-\mmb^2\rg^2}&\mbox{for
\mbb},\end{cases}\label{Vmab}\eeq
if we defined the ratios $r_{ij}=-\ld_i/\ld_j$ with $i,j=1,2$ and
identify $\ld$ and $M_i$ as follows
\beq\ld= \begin{cases}
%\begin{array}{rl}
\lda~~\mbox{and}~~\mma=M/\sqrt{\lda}&\mbox{for~\ma\ and \mba},\\
\ldb~~\mbox{and}~~\mmb=M/\sqrt{\ldb}&\mbox{for~\mbb}.\end{cases}\label{ldab}\eeq
From the first relation in \Eref{Vmab}, we easily infer that the
elimination of the pole from the denominator of $\Vhi$ can be
implemented if we set $N=2$ and take $\rs=\rss-1$ and $\mma$ to
tend to zero. On the other hand, no $N$ dependence in $\Vhi$
arises for \mb.

\begin{figure}[!t]\centering
\includegraphics[width=60mm,angle=-90]{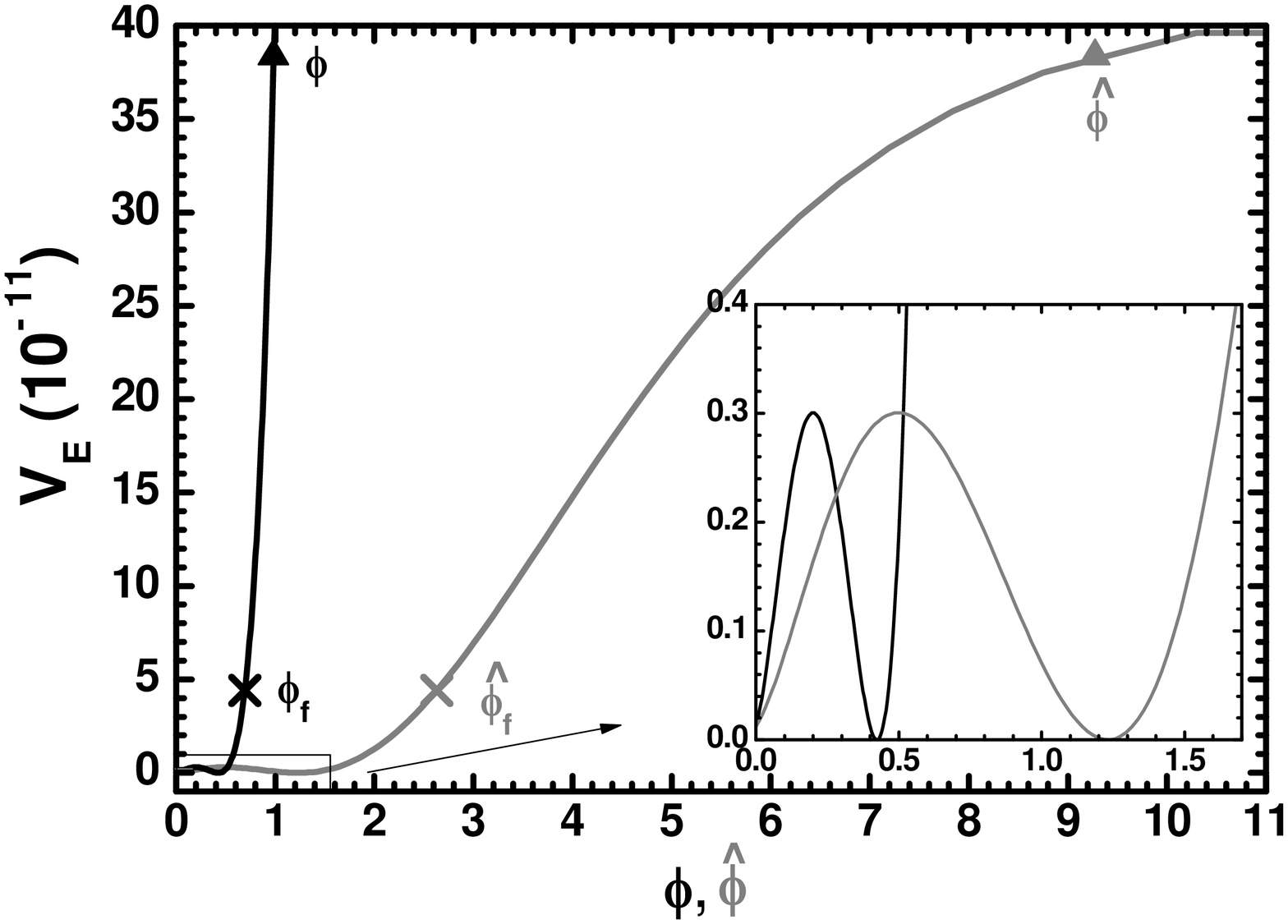}
\caption[]{\sl\small Inflationary potential $\Vhi$ for
\textsf{\ftn EM4} as a function of $\sg$ (black line) for $\sg>0$
and $\se$ (gray line) for $\se>0$. We set $\rrs=0.4$, $\mmb=0.1$
and $N=10$. Values corresponding to $\sgx$ [$\se$] and $\sgf$
[$\sef$] are depicted. Shown is also the low-$\sg$ behavior of
$\Vhi$ in the inset.}\label{fig0}
\end{figure}

The activation of the inflationary stage in our models is pretty
stable and well known \cite{alinde, sor}. It is based on the fact
that $\Vhi$ expressed in terms of $\se$ develops a plateau for
$\se\gg1$ but $\sg<1$ -- since $\se$ increases w.r.t $\sg$ as
inferred from \Eref{se}. This key feature is clearly demonstrated
in \Fref{fig0} where we comparatively plot $\Vhi$ for \mbb\ as a
function of $\sg$ (black line) and $\se$ (gray line) -- similar
behavior is expected for \ma\ and \mba. We use $N=10$, $\mmb=0.1$
and $\rrs=0.4$ resulting, via the inflationary constraints -- see
\Sref{res1} --, to $\ldb=3.5\cdot10^{-5}$. We see that $\Vhi$
experiences a stretching for $\se>1$ which results to a plateau
facilitating, thereby, the establishment of EMI. The two crucial
values of $\sg$ [$\se$] $\sgf=0.69$ [$\sef=2.63$] and $\sgx=0.984$
[$\sex=9.27$] which are defined in \Sref{res1} and limit the
observationally relevant inflationary period are also depicted.

%%%%%%%%%%%%%%%%%%%%%%%%%%%%%%%%%%%%%%%%%%%%%%%%%%%%%%%%%%%%%%%%%%%%
%%%%%%%%%%%%%%%%%%%%%%%%%%%%%%%%%%%%%%%%%

\subsection{\small\sf  SUSY Vacuum and Inflaton Mass}\label{sugra3}

Another important issue related to $\Vhi$ is the structure of the
SUSY vacuum. For \ma\ and \mba\ the extremization condition yields
the following critical points
\beq
\label{ext1}\sg_{1\pm}=\frac{1\pm\sqrt{1-4\rss\mma^2}}{2\rss}~~\mbox{and}~~\sg_{\rm
1c}=\frac1{2\rss},\eeq
from which we easily conclude that the SUSY vacuum lies at
$\vev{\sg}=\sg_{1-}$. Note that, for $|\rss|\leq1$, we obtain
$\sg_{1+}>1$ and so it is irrelevant for our scenario. To place
the location of the maximum $\sg_{\rm 1max}=\sg_{\rm 1c}$ at the
same invisible domain of $\sg$ values, we impose \beq \rss\leq1/2.
\label{rssb}\eeq As a consequence, EMI can take place undoubtedly
for $\sg$ values in the range $\vev{\sg}\leq\sg\leq1$. On the
other hand, the constraint $\vev{\sg}\leq1$ yields a maximal
$\mma$ value which is
\beq \label{mx1} M_{1\rm max}=\begin{cases}\sqrt{1-\rss},& \mbox{if}~~\rss<1/2,\\
1/2\rss& \mbox{if}~~\rss\geq1/2.\end{cases}\eeq
For \ma, $\rss\simeq1$ and so only the latter bound is applicable.
It is, though, overshadowed by the inflationary requirements which
dictate $\mma\lesssim0.01$ -- see \Sref{res}. For \mba, the bound
of \Eref{rssb} cancels the second branch in \Eref{mx1}.

\begin{figure}[!t]\centering
\includegraphics[width=60mm,angle=-90]{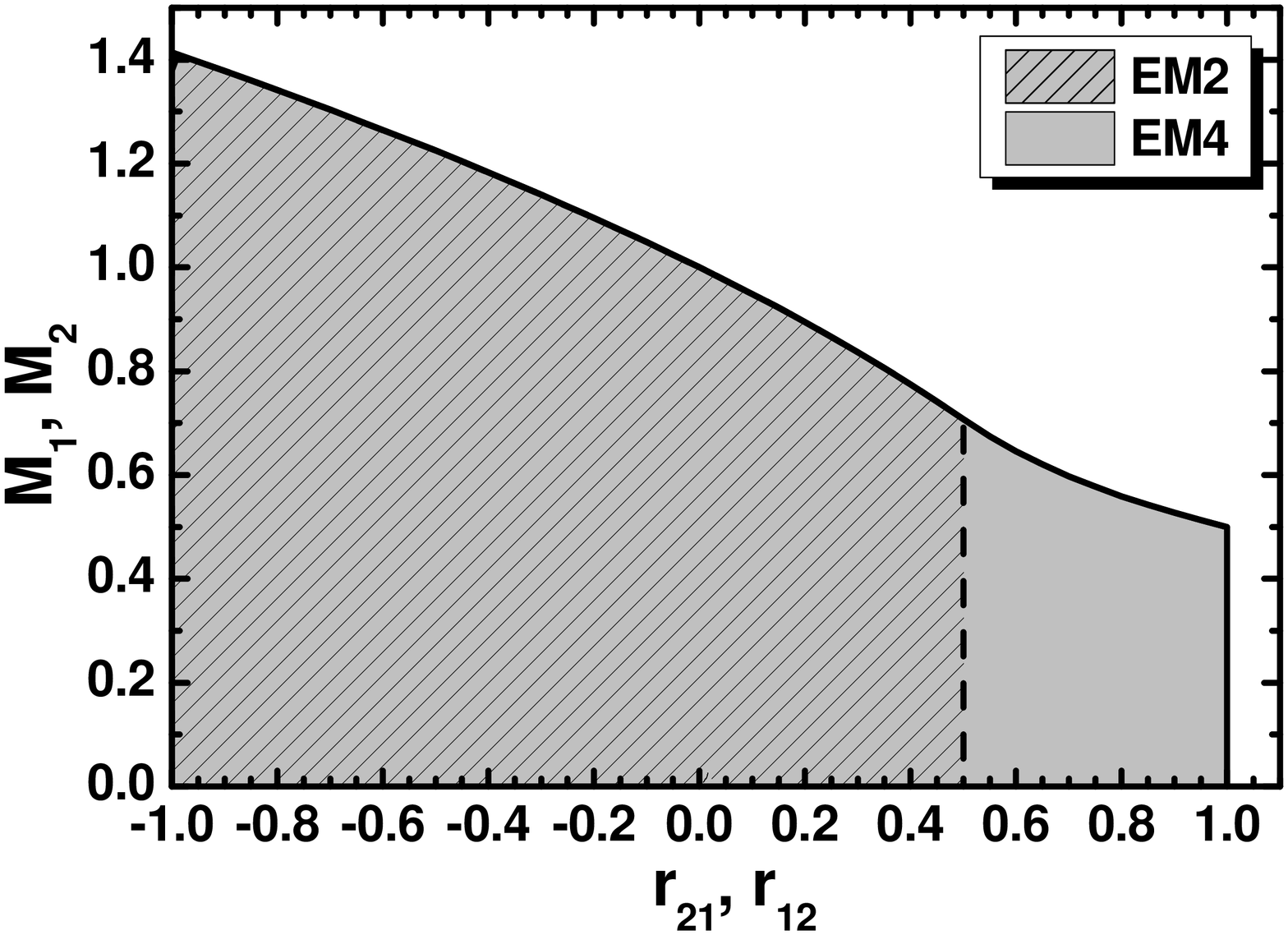}
\caption[]{\sl\small Allowed regions in the $\rss-\mma$ (lined
region) for \textsf{\ftn EM2} and $|\rss|\leq1$  and in the
$\rrs-\mmb$ (gray region) plane for \textsf{\ftn EM4} and
$|\rrs|\leq1$.}\label{fig00}
\end{figure}

A richer vacuum configuration arises for \mbb. Indeed, the extrema
of $\Vhi$ turn out to be
\beq
\label{ext2}\sg_{2\pm}=\frac12\lf\rrs\pm\sqrt{\rrs^2+4\mmb^2}\rg~~\mbox{and}~~\sg_{2\rm
c}=\frac{\rrs}{2},\eeq
from which the former are minima whereas the latter is maximum.
All of these are located below the pole position at $\sg=1$ for
the relevant $\rrs$ values. Therefore, the SUSY vacuum lies at one
of the two asymmetric $\vev{\sg}=\sg_{2\pm}$. The emergent
situation is depicted in the inset of \Fref{fig0} where we obtain
$\vev{\sg}=0.42$ or $-0.023$ and $\sg_{\rm max}=0.2$. The
constraint $\vev{\sg}\leq1$ yields
\beq \label{mx2}\mmb \leq M_{2\rm max}=\sqrt{1-\rrs}.\eeq
The restrictions in \eqs{mx1}{mx2} for \mba\ and \mbb\ are
depicted in \Fref{fig00} where we plot the allowed values in the
$\rss-\mma$ plane \mba\ and in $\rrs-\mmb$ plane for \mbb. We
confine ourselves to the realistic ranges $|\rss|\leq1$ and
$|\rrs|\leq1$ where inflationary solution are detected.

%%%%%%%%%%%%%%%%%%%%%%%%%%%%%%%%%%%%%%%%%%%%%%%%%%%%%%%%%%%%%%%%%%%%
%%%%%%%%%%%%%%%%%%%%%%%%%%%%%%%%%%%%%%%%%

The determination of $\vev{\sg}$ form  \eqs{mx1}{mx2} influences
heavily the mass of the (canonically normalized) inflaton,
\beqs\bal\dphi=\vev{J}\dph~~\mbox{with}~~\vev{J}=\sqrt{\frac{N}{2}}\vev{\fr}^{-1}
~~\mbox{\&}~~\dph=\phi-\vev{\phi} \label{dphi}
\end{align}
which is given by
\beq \label{msn} \msn=\left\langle\Ve_{\rm
E,\se\se}\right\rangle^{1/2}\simeq
\begin{cases}
{2\lda}/{\sqrt{N}}&\hspace*{-0.1cm}\mbox{for \ma\ \& \mba,}\\

{2\rrs\ldb}/{\sqrt{N}}&\hspace*{-0.1cm}\mbox{or}\\
\frac{2\rrs}{\sqrt{N}}(1-\rrs){\ldb}&\hspace*{-0.1cm}\mbox{for
\mbb,}\end{cases}\eeq\eeqs
where the last (approximate) equality is valid only for $M\ll1$.
Recall that we use $N=2$ for \ma\ and we obtain two possible vacua
for \mbb, as mentioned below \Eref{ext2}. For the parameters
employed in \Fref{fig0}, we obtain $\msn=10^{-5}\mP$ or
$5.6\cdot10^{-6}\mP$ corresponding to $\vev{\sg}=\sg_+$ and
$\sg_-$ respectively. The precise value of $\vev{\sg}$ requires a
numerical study of the inflaton dynamics after EMI which is
obviously beyond the scope of the present paper.

%However, for $\rrs\ll1$ and $\mmb\ll1$, the discrepancy between
%$\sg_+$ and $\sg_-$ is in practice negligible, whereas for higher
%values, $\vev{\sg}=\sg_+$ is more probable due to the maximum
%appearing during the evolution of $\sg$.

\subsection{\small\sf  Radiative Corrections}\label{sugra4}

To consolidate our proposal we verify that the configuration in
\Eref{inftr} is stable w.r.t the excitations of the non-inflaton
fields. Taking into the limit $\rs=\mma=0$ for \ma, $\rss=\mma=0$
for \mba\ and $\rrs=\mmb=0$ for \mbb, we find the expressions of
the masses squared $\what m^2_{\chi^\al}$ (with $\chi^\al=\th$ and
$s$) arranged in \Tref{tab1}. We there display the masses $\what
m^2_{\psi^\pm}$ of the corresponding fermions too -- we define
$\what\psi_{\Phi}=J\psi_{\Phi}$ where $\psi_\Phi$ and $\psi_S$ are
the Weyl spinors associated with $S$ and $\Phi$ respectively. We
notice that the relevant expressions can take a unified form for
all models -- recall that we use $N=2$ in \ma\ -- and approach,
close to $\sg=\sgx\simeq1$, rather well the quite lengthy, exact
ones employed in our numerical computation. From them we can
appreciate the role of $\nb<6$ in retaining positive $\what
m^2_{s}$. Also, we confirm that $\what
m^2_{\chi^\al}\gg\Hhi^2\simeq\Vhio/3$ for $\sgf\leq\sg\leq\sgx$.

Inserting the derived mass spectrum in the well-known
Coleman-Weinberg formula, we  can find the one-loop radiative
corrections, $\dV$ to $\Vhi$. Our inflationary results are immune
from $\dV$, provided that the renormalization group mass scale
$Q$, is determined by requiring $\dV(\sgx)=0$ or $\dV(\sgf)=0$ --
see the relevant detailed discussion in \cref{jhep}. These
conditions yield $Q\simeq(0.25-9)\cdot10^{-5}$ and eliminate any
possible dependence of our results on the choice of $Q$. Under
these circumstances, our results in the SUGRA set-up can be
exclusively reproduced by using $\Vhi$ in \Eref{Vmab}.

\begin{table}[t!]
\caption{\sl Mass spectrum along the path in \Eref{inftr} for EMI
-- \\ we take $n=1$ for \textsf{$\delta$\ftn EM} and \textsf{\ftn
EM2} whereas $n=2$ for \textsf{\ftn EM4}. }
\begin{ruledtabular}
\begin{tabular}{c|c|c|c}
%\begin{tabular}{c|@{\hspace{0.1cm}}c@{\hspace{0.1cm}}|@{\hspace{0.1cm}} c}\toprule
%
{\sc Fields}&{\sc Eingestates} {\hspace*{0.3cm}}&
\multicolumn{2}{c}{\sc Mass Squared}{\hspace*{0.3cm}}\\ \hline
%
%\multicolumn{3}{c}{Bosons}\\ \hline
%
$1$ real scalar &$\what \th${\hspace*{0.3cm}} & $\what m^2_{\th}${\hspace*{0.3cm}} &$6\Hhi^2${\hspace*{0.3cm}}\\
$2$ real scalars &$\what{s},\what{\bar s}${\hspace*{0.3cm}}  & $\what m^2_{s}$ {\hspace*{0.3cm}}&$6\Hhi^2/\nb${\hspace*{0.3cm}}\\
\hline\\[-0.4cm]
%
%\multicolumn{3}{c}{Fermions}\\ \hline
%
$2$ Weyl spinors&$({{\psi}_{S}\pm \what{\psi}_{\Phi})/\sqrt{2}}$
{\hspace*{0.3cm}}&
$\what{m}^2_{\psi\pm}${\hspace*{0.3cm}}&$6n\fr^2\Hhi^2/N\sg^2${\hspace*{0.3cm}}
%\\\botrule
\end{tabular}
\end{ruledtabular}\label{tab1}
\end{table}
%(hatted)
%

\section{Inflation Analysis}\label{res}

We here remind the basics of the inflation analysis in \Sref{res1}
and analyze the response of the proposed models analytically and
numerically in \Sref{ana} and \ref{res2} respectively.

\subsection{\small\sf  Preliminaries}\label{res1}

Let us recall that the investigation of slow-roll EMI requires the
computation of:\vspace*{-0.2cm}

\subparagraph{\sl (a)} The slow-roll parameters
\beqs \beq\label{sr}\epsilon= \left({\Ve_{\rm
E,\se}/\sqrt{2}\Ve_{\rm E}}\right)^2
\>\>\>\mbox{and}\>\>\>\>\>\eta={\Ve_{\rm E,\se\se}/\Ve_{\rm E}},
\eeq
which control the duration of EMI upon the condition
\beq\max\{\eph(\phi),|\ith(\phi)|\}\leq1.\label{srcon}\eeq \eeqs
Its saturation at $\sg=\sgf$ or $\se=\sef$ signals the termination
of EMI.

\subparagraph{\sl (b)} The number of e-foldings $\Ns$ that the
scale $\ks=0.05/{\rm Mpc}$ experiences during EMI and the
amplitude $\As$ of the power spectrum of the curvature
perturbations generated by $\sg$. These observables can be
calculated using the standard formulae
\begin{equation}
\label{Nhi}  \Ns=\int_{\sef}^{\sex} d\se\frac{\Vhi}{\Ve_{\rm
E,\se}}~~\mbox{and}~~\As= \left.\frac{1}{12\pi^2}\frac{\Ve_{\rm
E}^{3}}{\Ve^2_{\rm E,\se}}\right|_{\se=\sex},\eeq
where $\sgx~[\sex]$ is the value of $\sg~[\se]$ when $\ks$ crosses
the inflationary horizon.

\subparagraph{\sl (c)} The (scalar) spectral index, $\ns$, its
running, $\as$, and the tensor-to-scalar ratio, $r$, found from
the relations
\beqs\bea \label{ns} && \ns=\: 1-6\epsilon_\star\ +\
2\eta_\star,~~r=16\epsilon_\star, \\ \label{as} && \as
=\:2\left(4\eta_\star^2-(n_{\rm s}-1)^2\right)/3-2\xi_\star,
\eea\eeqs
where the variables with subscript $\star$ are evaluated at
$\sg=\sgx$ and $\xi={\Ve_{\rm E,\widehat\phi} \Ve_{\rm
E,\widehat\phi\widehat\phi\widehat\phi}/\Ve^2_{\rm E}}$.

\subsection{\small\sf  Analytic Results}\label{ana}

Applying the formulas of the section above we here present an
analytic approach to the dynamics of our models. Despite the fact
that the formulae is presented in terms of the field $\se$, we use
as independent variable $\sg$ taking advantage of \eqs{Vmab}{Je}
and avoiding the direct reference to $\Vhi=\Vhi(\se)$ which is
more complicated. In particular, \Sref{ana1}, \ref{ana2} and
\ref{ana3} are devoted to \ma, \mba\ and \mbb\ respectively.

\subsubsection{\small\sf  $\delta$ E Model (\ma)}\label{ana1}

For the present model we set $N=2$ -- see \Sref{sugra2}.  The
slow-roll parameters read
\beqs\begin{align}
\label{sr1e}\eph&\simeq\frac{2\fr^2}{\sg^2}-\frac4{\sg^3}\bigg({\rs\sg^2-2\mma^2(1-2\sg)}\bigg);\\
\nonumber\ith&\simeq2\lf2+\frac1{\sg^2}-\frac{3}{\sg}\rg+\frac2{\fr\sg^3}\bigg(\rs\sg^2(3\sg-4)\\&~~~~~~~~~~~~~~~
+\mma^2(2+\sg(4\sg-7))\bigg).\label{sr1i}\end{align}\eeqs
The condition of \Eref{srcon} in the present case implies
\beq \sg_{\rm f}\simeq \max \left\{2-\sqrt{2},
\sqrt{3}-1\right\},\label{sgf1}\eeq
i.e., $\sg_{\rm f}$ is determined due to the violation of the
$\ith$ criterion. Assuming $\sgf\ll\sgx$, $\Ns$ can be
approximately computed from \Eref{Nhi} as follows
\begin{equation}
\label{Nhi1}  \Ns\simeq
\frac{1}{2}\frac{\sgx}{1-\sgx}\>\>\>\Rightarrow\>\>\>\sgx\simeq
\frac{2\Ns}{2\Ns+1},
\end{equation}
where any dependence on the parameters $\rs$ and $M$ can be safely
ignored. From the last expression, we infer that $\sgx$ is
slightly lower than unity. Plugging $\sgx$ from \Eref{Nhi1} into
the rightmost equation in \Eref{Nhi} and solving w.r.t $\ld$, we
arrive at the expression
\beq \ld=2\sqrt{3\As}\pi/\Ns,\label{lan1}\eeq
which yields  values comparable to the ones obtained for the
quartic power-law model \cite{plin}. Upon substitution of $\sgx$
into \eqs{ns}{as}, we obtain the observational predictions of \ma\
which are
\beqs\bal  \label{ns1}\ns&\simeq
1-\frac{2}{\Ns}\lf1+\rs+{\mma^2\over2}\rg+4(\rs+\mma^2)(1-4\Ns);\\
r&\simeq\frac{8}{N_{\star}^2}\lf\frac{1-4(\rs+\mma^2)N_{\star}^2}{1-2(\rs+\mma^2)N_{\star}}\rg^2
\>\>\>\mbox{and}\>\>\>
\as\sim-\frac{3}{N_{\star}^2},\label{rs1}\end{align}\eeqs
where the last expression (for $\as$) yields just an order of
magnitude estimation. From \Eref{ns1} we expect that small
deviations of $\rs$ from zero are capable to generate sizable
deviations of $\ns$ from its value within original Starobinsky
inflation. Such a deviation is not possible for $r$.

\subsubsection{\small\sf  E Model 2 (\mba)}\label{ana2}

The slow-roll parameters are here estimated as follows
\beqs\begin{align}
\label{sr2e}\eph&\simeq\frac{4\fr^2}{N}\lf\frac{1-2\rss\sg}{\sg(1-\rss\sg)-M^2}\rg^2;\\
\ith&\simeq\frac{4\fr}{N}\frac{1-(2+3\rss(2-3\sg))\sg}{(\sg(1-\rss\sg)-M^2)^2},\label{sr2i}\end{align}\eeqs
with the $N$ dependence being explicitly displayed. The condition
of \Eref{srcon} yields
\beq \sg_{\rm f}\simeq\max\left\{2\frac{2-\sqrt{N}}{4-\sqrt{N}},
2\frac{3-\sqrt{1-N}}{8-N}\right\}.\label{sgf2}\eeq
For $N\sim{\cal O}(10)$, $\sgf$ turns out to be close to
$\sgx\sim1$. Therefore, an accurate analytic estimation of $\Ns$
may include both contributions as follows
\beq  \label{Nhi2} \Ns=I_N(\sgx)-I_N(\sgf),\eeq
where the function $I_{N}$, found performing the integration in
\Eref{Nhi}, takes the form
\bal  \nonumber
I_N(\sg)&=\big({N}/{8\fss^2}\big)\bigg(2\fss({1-\rss
-\mma^2})/\fr\\&+2(\fss+2\rss^2)\ln\fr-\ln(1-2\rss\sg)\bigg),
\label{iN1}\end{align}
with $\fss=1-2\rss$. I.e., as $\rss$ approaches its maximum,
$1/2$, in \Eref{rssb} -- $I_N(\sg)$ gets enhanced causing via
\Eref{Nhi2} a sizeable deviation of $\Ns$ from its result obtained
for a monomial $\Vhi$. In such a delicate situation, all three
terms in the right-hand side of \Eref{iN1} are equally important
and the analytic solution of $\Ns$ w.r.t $\sg$ is not doable. In
this domain, the numerical computation is our last resort.
Focusing on $\rss\ll0.5$, neglecting $I_N(\sgf)$ and the
logarithmic contributions from $I_N(\sgx)$ in \Eref{Nhi2}, we may
solve \Eref{Nhi2} w.r.t $\sgx$ with result
\beq
\sgx\simeq4\fss\Ns/\big(4\Ns\fss+N(1-\rss-\mma^2)\big).\label{sgx2}\eeq
Comparing with the numerical results we verify that the estimation
above is more and more accurate as $N, \rss$ and $\mma$ decrease
below $10, 0.5$ and $0.1$ respectively. Plugging $\sgx$ from
\Eref{sgx2} into the rightmost equation in \Eref{Nhi} and solving
w.r.t $\ld$ we find
\beq
\ld\simeq\sqrt{3N\As/2}(4\Ns+N)\pi/2N^2_\star.\label{lan2}\eeq
Inserting  \Eref{sgx2} into \eqs{ns}{as} and performing an
expansion in series for low $1/\Ns$ we obtain
\beqs\bal \label{ns2} \ns&\simeq
1-\frac2\Ns-\frac{N}{N_\star^2\fss^2}\lf1-\frac52\rss-3\rss^2-\frac12\mma^2\rg;\\
\nonumber r&\simeq
\frac{4N}{N_\star^2\fss^2}\bigg(1-\lf8-{N}\lf1+4\mma^2\rg\rg\rss/{2}\\&~~~~~~~~~~~~~~~~~~~+4\rss^2+N\mma^2/2\bigg);\label{r2}\\
\as&\simeq\frac{2}{N_\star^2}-\frac{N}{N_\star^3\fss^2}\lf\frac52+(6+7\mma^2)\rss-7\rss^2-\mma^2\rg.\label{as2}\end{align}\eeqs
For $\fss\simeq1$ (i.e. $\rss\simeq0$) the results above reduce to
the well-known predictions of $\alpha$-attractors -- recall that
our results can be directly compared with those in \cref{tmodel,
alinde} setting $N=3\alpha$.

\subsubsection{\small\sf  E Model 4 (\mbb)}\label{ana3}

Repeating the steps followed in the previous sections, we can
obtain the corresponding formulas for \mbb. In particular, the
slow-roll parameters read
\beqs\begin{align}
\label{sr3e}\eph&\simeq\frac{4\fr^2}{N}\lf\frac{2\sg-\rrs}{\sg^2-\rrs\sg-\mmb^2}\rg^2;\\
\nonumber\ith&\simeq\bigg({4\fr}/{N(\sg^4-\rrs\sg-\mmb^2)^2}\bigg)\bigg(2\sg^2(3-4\sg)
\\&-2\mmb^2(1-2\sg)+\rrs^2(1-2\sg)+3\rrs(3\sg-2)\sg\bigg).\label{sr3i}\end{align}\eeqs
The condition of \Eref{srcon} implies
\beq \sg_{\rm f}\simeq\mbox{\sf
max}\left\{4\frac{4-\sqrt{N}}{16-\sqrt{N}},
2\frac{14-\sqrt{2}\sqrt{2+3N}}{32-N}\right\}.\label{sgf3}\eeq
As in \Sref{ana2}, the reliable estimation of $\Ns$ requires the
application of \Eref{Nhi2} where the function $I_N$ is now given
by
\bal  \nonumber
I_N(\sg)&=\big({N}/{8\ffs^2}\big)\bigg(2\ffs({1-\rrs
-\mmb^2})/\fr\\&+2(\ffs-\rrs+\rrs^2)\ln\fr-\rrs^2\ln(2\sg-\rrs)\bigg)
\label{iN2}\end{align}
with $\ffs=2-\rrs$. Note that \Eref{mx2} protects the present case
from the appearance of any pole in the equation above. We may then
solve \Eref{Nhi2} w.r.t $\sgx$ with result
\beq
\sgx\simeq4\ffs\Ns/\big(4\ffs\Ns+(1-\rrs-\mmb^2)N\big),\label{sgx3}\eeq
where we neglected $I_N(\sgf)$ and the logarithmic contributions
from $I_N(\sgx)$. Plugging the expression above into the rightmost
equation in \Eref{Nhi} we find
\beq
\ld\simeq\sqrt{3N\As/2}(8\Ns+N)^2\pi/32N_\star^3.\label{lan3}\eeq
Moreover, substitution of \Eref{sgx3} into \eqs{ns}{as} yields
\beqs\bal \label{ns3} \ns&\simeq
1-\frac2\Ns-\frac{N}{N_\star^2\fss^2}\lf3-\frac52\rrs-\rrs^2+3\mmb^2\rg;\\
\nonumber r&\simeq
\frac{16N}{N_\star^2\ffs^2}\bigg(1-\lf1-{N}/8\Ns-\rrs/4\rg\rrs\\&~~~~~~~~~~~~~~~~~~~~+{N\mmb^2}\lf4-{\rrs}/{N}\rg/{8}\Ns\bigg);\label{r3}\\
\as&\simeq-\frac{8}{N_\star^3\fss^2}\lf1-\rrs+\rrs^2/4+7N/8\Ns\rg.\label{as3}\end{align}\eeqs
Again, for $\ffs\simeq2$ (i.e. $\rrs\simeq0$) the results above
reduce to the well-known predictions of $\alpha$-attractors
\cite{tmodel, alinde}.

\subsection{\small\sf Numerical Results}\label{res2}

Our analytic findings above can be checked and extended for any
value of the relevant parameters numerically. At first, we
confront the quantities in \Eref{Nhi} with the observational
requirements \cite{plcp}
\bal& \Ns \simeq61.3+\frac{1-3w_{\rm rh}}{12(1+w_{\rm
rh})}\ln\frac{\pi^2g_{\rm rh*}\Trh^4}{30\Vhi(\sgf)}+\nonumber\\
&\frac14\ln{\Vhi(\sgx)^2\over g_{\rm
rh*}^{1/3}\Vhi(\sgf)}~~\mbox{and}~~\As\simeq2.105\cdot10^{-9},~~~~~~\label{Prob}\end{align}
%}{
where we assume that EMI is followed in turn by an oscillatory
phase with mean equation-of-state parameter $w_{\rm rh}$,
radiation and matter domination. Motivated by implementations
\cite{univ} of non-thermal leptogenesis, which may follow EMI, we
set $\Trh\simeq10^9~\GeV$ for the reheat temperature. Also, we
take for the energy-density effective number of degrees of freedom
$g_{\rm rh*}=228.75$ which corresponds to the MSSM spectrum. Note
that this $\Trh$ avoids exhaustive tuning on the relevant coupling
constant involved in the decay width of the inflaton -- cf.
\cref{ellis21}.

Due to the polynomial character of $\Vhi$ in \Eref{Vmab} and the
non-minimal kinetic mixing in \Eref{Je}, the estimation of $\wrh$
requires some care -- cf. \cref{wreh}. We determine it adapting
the general formula \cite{turner}, i.e.
\beq w_{\rm rh}=2\frac{\int_{\sgn}^{\sgm} d\sg J(1-
\Vhi/\Vhi(\sgm))^{1/2}}{\int_{\sgn}^{\sgm} d\sg J(1-
\Vhi/\Vhi(\sgm))^{-1/2}}-1,\label{wrh}\eeq
where $\sgn=\vev{\sg}$ is given by \Eref{ext1} for \ma\ and \mba\
or \Eref{ext2} for \mbb -- in the latter case we assume that
$\vev{\sg}=\sg_{+}$ which is considered as more probable than the
$\vev{\sg}=\sg_{-}$. The amplitude of the oscillations during
reheating $\sgm$ is found by solving numerically the condition
$\sqrt{3}\Hhi(\sgm)=\mq$ if $\sqrt{3}\Hhi(\sgf)>\mq$ or it is
$\sgm=\sgf$ otherwise. Performing an expansion for low $\sgm$
values we can obtain the following approximate expression
\beqs\bal w_{\rm rh}\simeq
-\frac{2\sgm}{3\pi}+\lf\frac{4}{3\pi^2}-\frac14\rg\sgm^2+\frac{\pi^2-80}{30\pi^3}\sgm^3\label{wrh1}\end{align}
for \ma\ where $\rss\simeq1$. For \mba, where $\rss\neq1$ the
corresponding expression takes the form
\bal \nonumber w_{\rm
rh}&\simeq-\frac{2\sgm}{3\pi}(1+2\rss)\\&+\sgm^2\lf\frac{4}{3\pi^2}+\frac{16\rss}{3\pi^2}+
\frac{2\rss}{3\pi}-\frac14(1+3\rss)\rg\,.
\label{wrh2}\end{align}\eeqs
It was not doable to obtain similar formula for \mbb, for which we
trust our numerical result. These refinements result to a
reduction of $\wrh$ w.r.t its value for a pure quadratic
($\wrh=0$) or quartic ($\wrh=1/3$) potential.

Enforcing \Eref{Prob} we can restrict $\ld$ and $\sgx$ via
\Eref{Nhi}. In general, we obtain $\ld\simeq(0.1-5.4)\cdot10^{-5}$
in agreement with \eqss{lan1}{lan2}{lan3}. Regarding $\sgx$ we
assume that $\sg$ starts its slow roll below the location of
kinetic pole, i.e., $\sg=1$, consistently with our approach to
SUGRA as an effective theory below $\mP=1$. The closer to pole
$\sgx$ is set the larger $\Ns$ is obtained. Therefore, a tuning of
the initial conditions is required which can be somehow quantified
defining the quantity
\beq \Dex=\left(1 - \sgx\right).\label{dex}\eeq
The naturalness of the attainment of EMI increases with $\Dex$.
After the extraction of $\ld$ and $\sgx$, we compute the models'
predictions via \eqs{ns}{as}, for any selected values for the
remaining parameters. Our outputs are encoded as lines in the
$\ns-r$ plane and compared against the observational data
\cite{gws1} in Figs.~\ref{fig1} for \ma\ and \ref{fig2} for \mba\
and \mbb. We take into account the latest \emph{Planck release 4}
({\sf\ftn PR4}) -- including TT,TE,EE+lowlEB power spectra
\cite{pl20} --, \emph{Baryon Acoustic Oscillations} ({\sf\ftn
BAO}), CMB-lensing and BICEP/{\it Keck} ({\sf\ftn BK18}) data
\cite{gws1}. Fitting it \cite{gws} with $\Lambda$CDM we obtain the
marginalized joint $68\%$ [$95\%$] regions depicted by the dark
[light] shaded contours in the aforementioned figures.
Approximately we obtain
\beq \label{nswmap} \ns=0.965\pm0.009~~\mbox{and}~~r\lesssim0.032,
\eeq at 95$\%$ \emph{confidence level} ({\sf\ftn c.l.}) with
negligible $\as$. The results are exposed separately in
\Sref{res2a} for \ma\ and \ref{res2b} for \mb.

%%%%%%%%%%%%%%%%%%%%%%%%%%%%%%%%%%%%%%%%%%%%%%%%%%%%%%%%%%%%%%%%%%%%
\begin{figure}[!t]
\centering
\includegraphics[width=60mm,angle=-90]{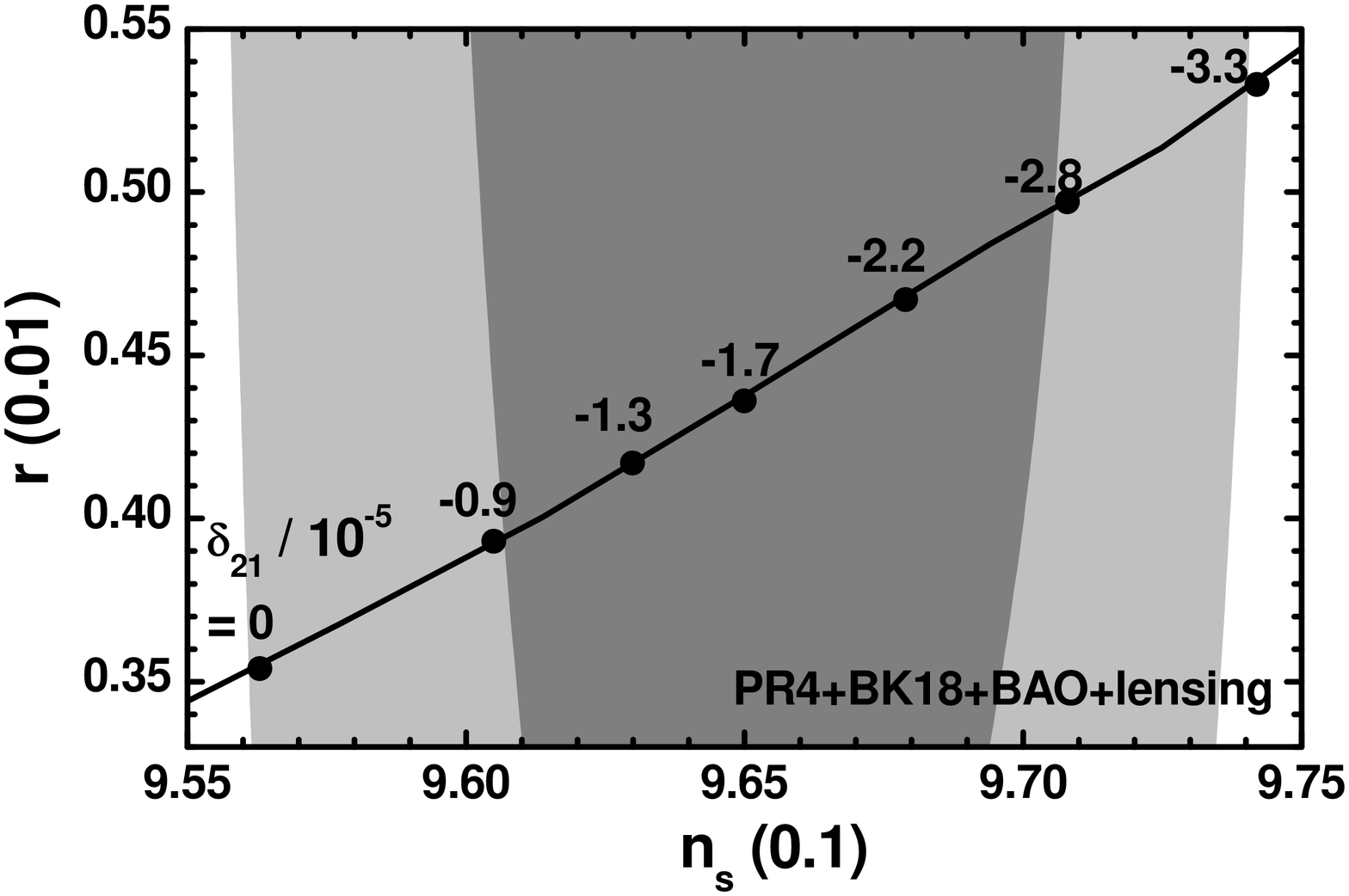}\\
\begin{ruledtabular}
\begin{tabular}{c||c|c|c}%\toprule
$(\ns/0.1,r/0.01)$&\multicolumn{1}{c|}{$\mma=0.001$}&\multicolumn{1}{c|}{$\mma=0.005$}&\multicolumn{1}{c}{$\mma=0.01$}\\
\cline{2-4}& $-\rs/10^{-5}$&$-\rs/10^{-5}$&$-\rs/10^{-4}$
\\\colrule
%,,,,
$(9.56, 0.35)$&$0$&$2.45$&$0.98$\\
$(9.61, 0.39)$&$0.9$&$3.25$&$1.08$\\
$(9.65, 0.44)$&$1.7$&$4.1$&$1.15$\\
$(9.71, 0.5)$&$2.8$&$5.2$&$1.26$\\
$(9.74, 0.53)$&$3.3$&$5.7$&$1.315$\\%\botrule
\end{tabular}
\end{ruledtabular}
\caption{\sl \small Allowed curve in the $\ns-r$ plane for
\textsf{\ftn $\delta$EM} with $\mma=0.001$ and various $\rs$'s
indicated on the line. The marginalized joint $68\%$ [$95\%$] c.l.
regions \cite{gws} from PR4, {\sffamily\ftn BK18}, BAO and lensing
data-sets \cite{gws1} are depicted by the dark [light] shaded
contours. The allowed curve remains essentially unaltered if we
set $\mma=0.005$ or $0.01$. The variation some of the $\rs$ values
indicated in the plot is shown in the table.}\label{fig1}
\end{figure}

%%%%%%%%%%%%%%%%%%%%%%%%%%%%%%%%%%%%%%%%%%%%%%%%%%%%%%%%%%%%%555

Let us note here, that urged by similar investigations
\cite{ellis21}, we performed a numerical comparison of our results
with those obtained if we use the second-order slow-roll
formulation \cite{rocky}. We verified that $\what
N_\star-\Ns\simeq1-2$ where $\what N_\star$ is the corrected $\Ns$
value. This refinement, though, has negligible impact on $\ns,
\as$ and $r$ and so it does not affect our results to any
essential way.

%%%%%%%%%%%%%%%%%%%%%%%%%%%%%%%%%%%%%%%%

\begin{figure*}[!t]
\includegraphics[width=60mm,angle=-90]{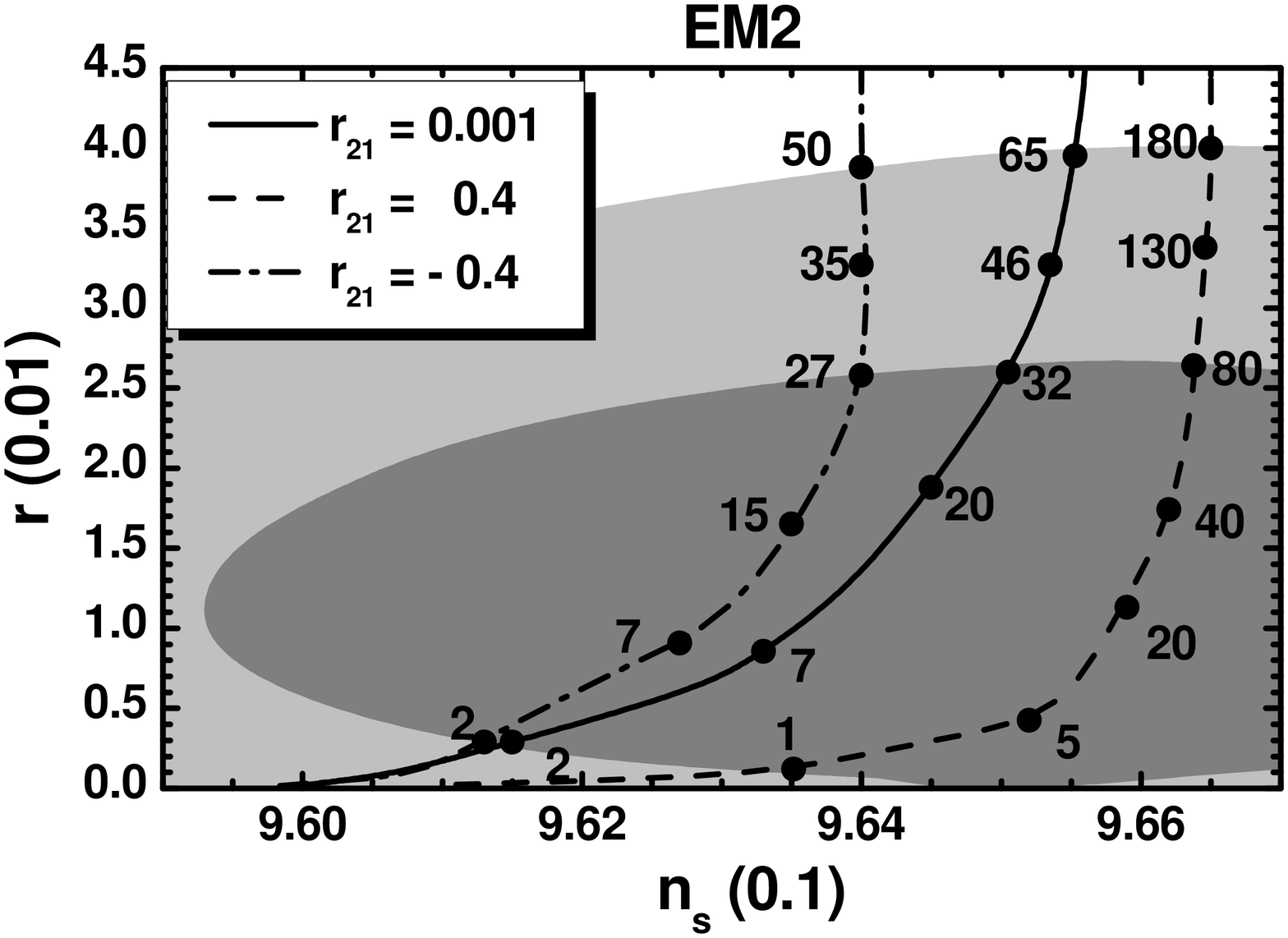}
\includegraphics[width=60mm,angle=-90]{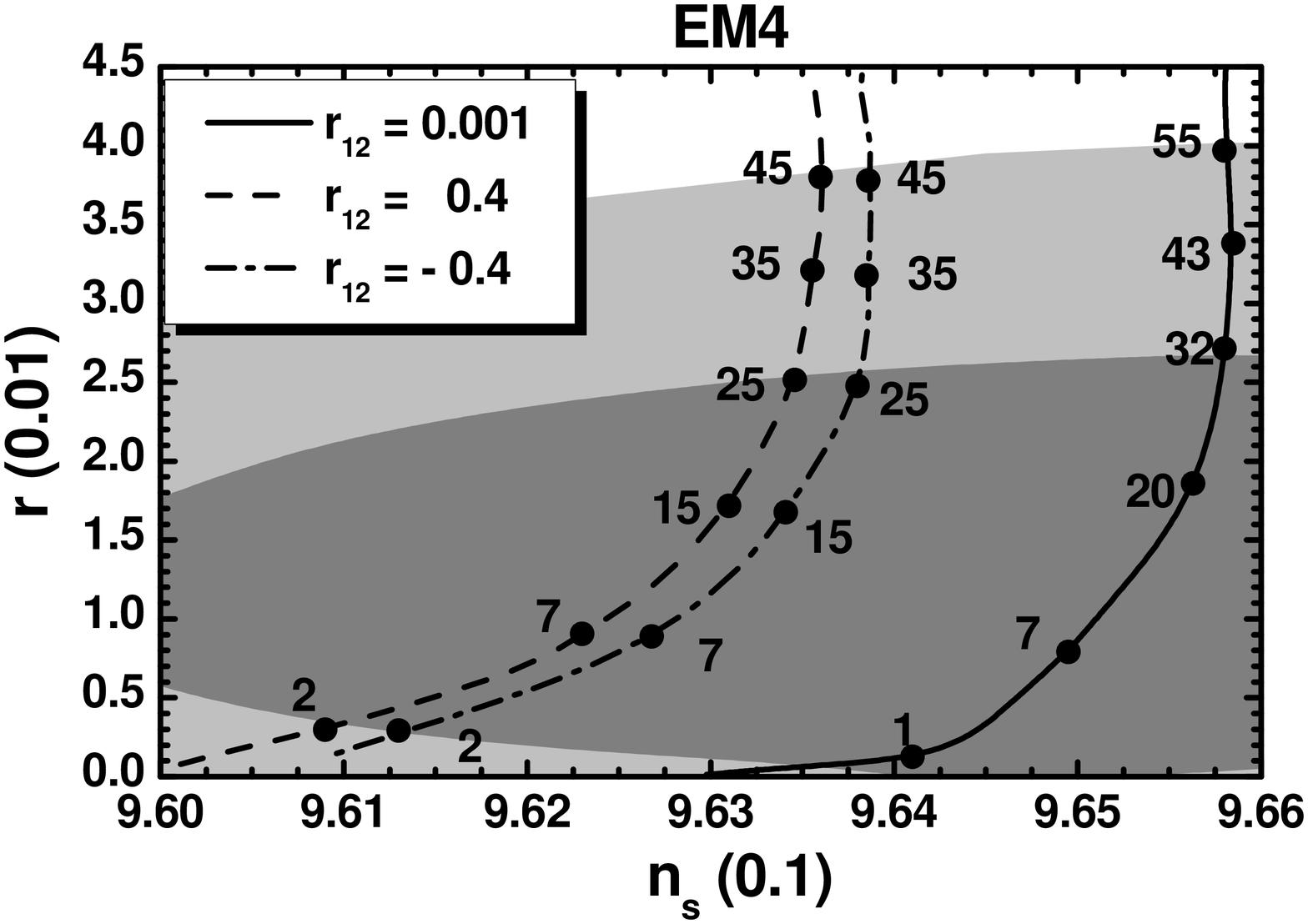}
\caption{\sl Allowed curves in the $\ns-r$ plane for \textsf{\ftn
EM2}, $\mma=0.01$, and $\rss=0.001$ (solid line), $\rss=0.4$
(dashed line), $\rss=-0.4$ (dot-dashed line) or \textsf{\ftn EM4},
$\mmb=0.01$, and $\rrs=0.001$ (solid lines), $\rrs=0.4$ (dashed
line), $\rrs=-0.4$ (dot-dashed line). The points indicated on the
curves correspond to various $N$ values. The shaded corridors are
identified as in \Fref{fig1}.}\label{fig2}
\end{figure*}
%%%%%%%%%%%%%%%%%%%%%%%%%%%%%%%%%%%%%%%%%%%%%%%%%%%%%%%%%%%%%%%%%

\subsubsection{\small\sf $\delta$ E Model (\ma)}\label{res2a}

Let, initially, recall that the remaining free parameters of \ma\
are just two, $\rs$ and $\mma$. As explained in \Sref{sugra2}, the
cancellation of $\fr$ from the denominator of $\Vhi$ in
\Eref{Vmab} requires a synergy between the conditions $N=2$,
$\mma\ll1$ and a tuning on $\rss$ -- cf.
\cref{eno5,eno7,nsreview,sor}. This unavoidable tuning is
qualified through the parameter $\rs=\rss-1$ which is varied along
the curve of \Fref{fig1} fixing $\mma=0.001$. We notice that $\rs$
is stuck at the level $10^{-5}$. Practically this result is intact
for any lower $\mma$ value. Varying $\mma$ for $0\leq\mma\leq0.02$
we can obtain a line undistinguished from that shown in
\Fref{fig1} with the depicted points $(\ns,r)$ corresponding to
readjusted $\rs$ values, though. For some sample $\mma$ values,
$\mma=0.005$ and $0.01$, the outputs of this readjustment are
listed in the Table in \Fref{fig1}. We see that increasing $\mma$
the tuning on $\rs$ is slightly ameliorated. The increase of
$\mma$, however, is severely limited from the viability of the EMI
setting. The resulting $\ns$ and $r$ increase with $\mma$ and
$\rs$. This increase, though, is more drastic for $\ns$ which
covers the whole allowed range in \Eref{nswmap}. From the
considered data we collect the results
\beqs\bal \label{resm1}
&-1.315\cdot10^{2}\lesssim\rs/10^{-6}\lesssim0
,~~9\cdot10^{-3}\lesssim\Dex\lesssim0.01,\\
\label{resm1a} &
0.9\lesssim-\as/10^{-3}\lesssim1.4~~\mbox{and}~~3.6\lesssim\msn/10^{13}\GeV\lesssim5
\end{align}\eeqs
with $(\ns, r)$ varying in the range indicated in the Table of
\Fref{fig1}. Here and hereafter, we restore for convenience units
regarding the results on $\msn$. In all cases, we obtain
$\Ns\simeq44$ consistently with \Eref{Prob} and the resulting
$\wrh\simeq-0.237$ from \Eref{wrh1}. The achieved $r$ values are
of the order of $10^{-3}$ and may be testable by the
next-generation of polarized CMB space missions, like LiteBIRD
\cite{bird}.

\subsubsection{\small\sf E Models 2 and 4}\label{res2b}

Turning now to \mb, we remind that the remaining free parameters
are $N$, and $\rss$ and $\mma$ for \mba\ or $\rrs$ and $\mmb$ for
\mbb\ -- see \Eref{Vmab}. Our results for these models are
presented in \Fref{fig2} and \Tref{tab2}. In the figure above, we
present the allowed curves in the $\ns-r$ plane varying $N$ along
them and fixing the other parameters to some representative
values.

Focusing first on the left plot of \Fref{fig2}, devoted to \mba,
we see that the solid, dashed, and dot-dashed lines are plotted
for $\rss=0.001$, $0.4$ and $-0.4$ correspondingly and
$\mma=0.01$. Along these lines $\Ns\simeq50$ with
$\wrh\simeq-(0.05-0.1)$ from \Eref{wrh2} independently from the
parameters $\mma$ and $\rss$. As a consequence \cite{turner},
$\Vhi$ for $\sg$ close to $\vev{\sg}$ behaves largely as a
quadratic potential. We remark that the results for $\rss=0.001$
and $-0.4$ yield $\ns$ a little lower than its central value in
\Eref{nswmap}. Interestingly enough, as $\rss$ approaches its
upper bound in \Eref{rssb} -- and $\fss$ in the denominator of the
function $I_N$ in \Eref{iN1} approaches zero --, $\ns$ increases
slightly and becomes precisely equal to its central value in
\Eref{nswmap} -- see the dashed line corresponding to $\rss=0.4$
in the left plot of \Fref{fig2}. In the same regime we also remark
that the upper bound on $N$ from \Eref{nswmap} is sharply
increased enhancing $\Dex$ too. Unfortunately, the analytic
formulas in \eqs{ns2}{r2} fail to explain further this observation
since these are not valid in this regime. The variation of our
results increasing $\mma$ beyond $0.01$, can be inferred by the
three leftmost columns of \Tref{tab2}, where we fix $N=10$ and
$\mma=0.5$ and list the outputs for the three $\rrs$ values used
in the left plot of \Fref{fig2}. We notice that our conclusions
regarding the increase of $\ns$ and $\Dex$ as $\rss$ approaches
$0.5$ insist. From the relevant data, we find the following
allowed ranges of parameters
\beqs\bal \label{resm2} &0.1\leq
N\leq180,~~0.961\lesssim\ns\lesssim0.966,~~0.00015\lesssim
r\lesssim0.4, \\ \label{resm2a}
&0.014\lesssim\Dex/0.1\lesssim3.4~~\mbox{and}~~1.9\lesssim\msn/10^{13}\GeV\lesssim5.6,
\end{align}\eeqs
where we set an artificial lower bound on $N$, $0.1$ to evade
extensive (theoretical) tuning. The maximal $\Dex$ and $\ns$
values are encountered for $\rss=0.4$, whereas $\as$ varies in the
range $-(6.6-7.7)\cdot10^{-4}$.

%%%%%%%%%%%%%%%%%%%%%%%%%%%%%%%%%%%%%%%%%%%%%%%%%%%%%%%%%%%%%%%%%%%
\begin{table}[b!]
\caption{\sl Parameters and observables for $N=10$, \textsf{\ftn
EM2} and $\mma=0.5$ or \textsf{\ftn EM4} and $\mmb=0.5$.}
\begin{ruledtabular}
\begin{tabular}{c||c|c|c|c|c|c}
%\toprule
{Model:} &\multicolumn{3}{c|}{{\sf EM2} with $\rss$ values:}
&\multicolumn{3}{c}{{\sf EM4} with $\rrs$ values:} \\\cline{2-7}
&$0.001$&{$0.4$}&$-0.4$&$0.001$&{$0.4$}&$-0.4$\\
\hline\hline
$\sgx/0.1$&$9.686$&$9.5$&$9.73$&$9.84$&$9.9$&$9.795$\\
$\Dex (\%)$&$3.1$&$5$&$2.7$&$1.6$&$0.9$&$2$\\
$\sgf/0.1$&{$5.4$}&$5.3$&$5.4$&$7.1$&$8.4$&$6.2$\\\hline
$\wrh$&$-0.08$&$-0.11$&$-0.06$&$-0.04$&$-0.06$&$-0.029$\\\hline
$\Ns$&$49.2$&$49.7$&$50.5$&$50.3$&$49.8$&$50.5$\\\hline
$\ld/10^{-5}$&$2.7$&$4.7$&$1.7$&$2.7$&$5.8$&$1.8$\\\hline
$\ns/0.1$&$9.62$&$9.65$&$9.62$&$9.62$&$9.63$&$9.62$\\
$-\as/10^{-4}$&$7.4$ &$6.8$&$7.2$&$7.2$&$7.1$&$7.2$\\
$r/10^{-2}$&$1.2$&$0.8$&$1.2$&$1.2$&$1.2$&$1.2$\\%\botrule
\end{tabular}\label{tab2}
\end{ruledtabular}
\end{table}

%%%%%%%%%%%%%%%%%%%%%%%%%%%%%%%%%%%%%%%%%%%%%%%%%%%%%%%%%%%%%%5

We proceed now with the investigation of \mbb. Fixing $\mmb=0.01$
we present the allowed solid, dashed and dot-dashed curves in the
right plot of \Fref{fig2} corresponding to $\rrs=0.001$, $0.4$ and
$-0.4$ respectively. In this case, the $\wrh$ values fluctuates
widely -- depending on $\rrs$ and $\mmb$ -- which influences $\Ns$
via \Eref{Prob}. Namely, for $\mmb\lesssim0.01$ and
$\rrs\lesssim0.1$ we obtain $\wrh\simeq(0.25-0.39)$ which results
to $\Ns\simeq(54-56)$ whereas for larger $\mmb$ and $\rrs$ values
we obtain $\wrh\simeq(0.0-0.086)$ which yields $\Ns\simeq(50-53)$.
I.e., in the former case, $\Vhi$ for $\sg\ll1$ behaves as a
quadratic potential whereas for the latter it approaches the
quartic potential \cite{turner}. The augmentation of $\Ns$ for low
$\rrs$ (and $\mmb$) values increases $\ns$ which converges towards
the ``sweet'' spot of the recent data in \Eref{nswmap}, as shown
in the right plot of \Fref{fig2}. Indeed, the solid line,
corresponding to $\rrs=0.001$ and $\mmb=0.01$, for $N\gtrsim7$
passes from $\ns$ values close to the central one in
\Eref{nswmap}. The variation of our results increasing $\mmb$
beyond $0.01$, can be inferred by the three rightmost columns of
\Tref{tab2}, where we fix $N=10$ and $\mmb=0.5$ and list the
outputs for the three $\rrs$ values used in the right plot of
\Fref{fig2}. Since $\mmb$ is larger than $0.1$ we obtain
$\wrh\simeq-0.05$ -- i.e. a quadratic-like potential --, and
$\Ns\simeq50$ and so $\ns\simeq0.963$. Comparing with the data of
the plot we conclude that increasing $\mmb$, the required $\sgx$
slightly increases lowering the tuning. In total, for \mbb\ we
obtain
\beqs\bal \label{resm3}
&0.1\leq N\leq55,~~0.961\lesssim\ns\lesssim0.965,~~0.00013\lesssim r\lesssim0.4, \\
\label{resm3a}
&0.1\lesssim\Dex/0.01\lesssim8~~\mbox{and}~~1.2\lesssim\msn/10^{13}\GeV\lesssim2.2\,.
\end{align}\eeqs
The minimal values of $\ns$ and $r$ are $0.961$ and $0.0007$,
respectively, encountered for the minimal $N$ value considered,
$0.1$ whereas $\as\simeq-0.0007$ constantly. As regards $\Dex$, we
see that tuning becomes milder w.r.t \ma\ but worse w.r.t that in
\mba. In general it is a little improved compared to the models in
\cref{sor}.

From the results above, we clearly infer that the polynomial form
of $W$ in \Eref{whi} has observational consequences which render
our models explicitly distinguishable from the usually adopted
monomial EMI \cite{plin, linde21, ellis21}.

\section{Conclusions}\label{con}

Embarking on the fact that EMI (i.e., E-model inflation) is
realized by the potential shown in \Eref{etm}, we introduced the
\Ka s in \eqs{Ks}{tkas} which naturally yield the required
$\sg(\se)$ function in \Eref{se0} connecting the initial, $\sg$,
with the canonically normalized, $\se$, inflaton. We then explored
the hyperbolic geometry of the adopted $K$'s and showed that they
are invariant under the set of transformations in \Eref{trn}
composing a set of matrices {\bf M} in \Eref{mts} which can be
related to the group $U(1,1)$ -- see \Eref{ms}.  Despite the fact
that their \Km\ does not enjoy a widely recognized symmetry, the
proposed $K$'s can coexist with the superpotential $W$ in
\Eref{whi} which is quite generic, renormalizable and consistent
with an $R$ symmetry. The emerging scalar potential gives rise to
three inflationary models (\ma, \mb) with potentials given in
\Eref{Vmab}. Within \ma\ the denominator of the potential is
cancelled out for $N=2$, low $M$ values and $\ldb\simeq-\lda$
whereas the lack of denominator in \mba\ and \mbb\ frees $N$,
$\lda$, $\ldb$ and $M$.

All the models excellently match the observations by restricting
the free parameters to reasonably ample regions of values. In
particular, within \ma\ any observationally acceptable $\ns$ is
attainable by tuning $\rs$ to values of the order $10^{-5}$,
whereas $r$ is kept at the level of $10^{-3}$ -- see \Eref{resm1}.
On the other hand, \mb\ avoid any tuning, larger $r$'s are
achievable as $N$ increases beyond $2$, while $\ns$ lies close to
its central value -- see \Eref{resm2}. For both latter models, we
localized portions of the allowed parameter space with $\ns$
precisely equal to its central observational value. The inflaton
mass is collectively confined to the range
$(1.2-5.6)\cdot10^{13}~\GeV$.

Comparing our setting with those \cite{tkref, tkin} which employ
the half-plane parameterization of the $SU(1,1)/U(1)$ \Km\ and
display also a kinetic pole of order one placed at zero, we should
note that the translation of the pole at unity within our
framework, although trivial at a first slight, leads to the
following far-reaching consequences: {\sf\ftn (i)} It changes the
geometry of the internal space; {\sf\ftn (ii)} it allows $W$ to
assume a simple and generic form controlled by an $R$ symmetry;
{\sf\ftn (iii)} it offers a more sizable variation of the
observables due to the extra free parameters. For these reasons,
we believe that the predictive, economical and
observationally-friendly EMI can be comfortably implemented within
the framework described in this paper.

\paragraph*{\small \bf\scshape Acknowledgments} {\small I would like to thank  S.~Ketov and E.W.
Kolb for interesting discussions. This research work was supported
by the Hellenic Foundation for Research and Innovation (H.F.R.I.)
under the ``First Call for H.F.R.I. Research Projects to support
Faculty members and Researchers and the procurement of high-cost
research equipment grant'' (Project Number: 2251).}

%\onecolumngrid \bec\rule{0.5\textwidth}{1.pt}\eec\vspace*{-1.2cm}

\def\ijmp#1#2#3{{\sl Int. Jour. Mod. Phys.}
{\bf #1},~#3~(#2)}
\def\plb#1#2#3{{\sl Phys. Lett. B }{\bf #1}, #3 (#2)}
\def\prl#1#2#3{{\sl Phys. Rev. Lett.}
{\bf #1},~#3~(#2)}
\def\rmp#1#2#3{{Rev. Mod. Phys.}
{\bf #1},~#3~(#2)}
\def\prep#1#2#3{{\sl Phys. Rep. }{\bf #1}, #3 (#2)}
\def\prd#1#2#3{{\sl Phys. Rev. D }{\bf #1}, #3 (#2)}
\def\npb#1#2#3{{\sl Nucl. Phys. }{\bf B#1}, #3 (#2)}
\def\npps#1#2#3{{Nucl. Phys. B (Proc. Sup.)}
{\bf #1},~#3~(#2)}
\def\mpl#1#2#3{{Mod. Phys. Lett.}
{\bf #1},~#3~(#2)}
\def\jetp#1#2#3{{JETP Lett. }{\bf #1}, #3 (#2)}
\def\app#1#2#3{{Acta Phys. Polon.}
{\bf #1},~#3~(#2)}
\def\ptp#1#2#3{{Prog. Theor. Phys.}
{\bf #1},~#3~(#2)}
\def\n#1#2#3{{Nature }{\bf #1},~#3~(#2)}
\def\apj#1#2#3{{Astrophys. J.}
{\bf #1},~#3~(#2)}
\def\mnras#1#2#3{{MNRAS }{\bf #1},~#3~(#2)}
\def\grg#1#2#3{{Gen. Rel. Grav.}
{\bf #1},~#3~(#2)}
\def\s#1#2#3{{Science }{\bf #1},~#3~(#2)}
\def\ibid#1#2#3{{\it ibid. }{\bf #1},~#3~(#2)}
\def\cpc#1#2#3{{Comput. Phys. Commun.}
{\bf #1},~#3~(#2)}
\def\astp#1#2#3{{Astropart. Phys.}
{\bf #1},~#3~(#2)}
\def\epjc#1#2#3{{Eur. Phys. J. C}
{\bf #1},~#3~(#2)}
\def\jhep#1#2#3{{\sl J. High Energy Phys.}
{\bf #1}, #3 (#2)}
\newcommand\jcap[3]{{\sl J.\ Cosmol.\ Astropart.\ Phys.\ }{\bf #1}, #3 (#2)}
\newcommand\njp[3]{{\sl New.\ J.\ Phys.\ }{\bf #1}, #3 (#2)}
\newcommand\jcapn[4]{{\sl J.\ Cosmol.\ Astropart.\ Phys.\ }{\bf #1}, no. #4, #3 (#2)}


\begin{thebibliography}{99}
 \section*{\refname}  %\ignorespaces


\bibitem{terada} T.~Terada, {\it Generalized Pole Inflation: Hilltop,
Natural, and Chaotic Inflationary Attractors,}
\plb{760}{2016}{674} [\arxiv{1602.07867}]; B.J.~Broy, M.~Galante,
D.~Roest and A.~Westphal, {\it Pole inflation, Shift symmetry and
universal corrections,} \jhep{12}{2015}{149} [\arxiv{1507.02277}].


\bibitem{pole} T.~Kobayashi, O.~Seto and T.H.~Tatsuishi, {\it Toward pole
inflation and attractors in supergravity: Chiral matter field
inflation}, {\sl Prog. Theor. Phys.}~\textbf{2017}, no.~12, 123B04
(2017) [\arxiv{1703.09960}].

\bibitem{sor} C.~Pallis, {\it Pole-Induced Higgs Inflation With Hyperbolic
Kaehler Geometries,} \jcap{05}{2021}{043} [\arxiv{2103.05534}];
C.~Pallis, {\it $SU(2,1)/(SU(2) \times U(1))$ B-L Higgs
Inflation,} {\sl J. Phys. Conf. Ser. }\textbf{2105}, no.~12, 12
(2021) [\arxiv{2109.06618}].
%%CITATION = doi:10.1088/1475-7516/2021/05/043%%
%%CITATION = doi:10.1088/1742-6596/2105/1/012007%%

\bibitem{new} S.~Karamitsos and A.~Strumia,
{\it Pole inflation from non-minimal coupling to gravity,}
\arxiv{2109.10367}; B.~Afshar, N.~Riazi and H.~Moradpour, {\it
Pole inflation in dRGT theory,} \arxiv{2110.02278}.


\bibitem{eno5} J.~Ellis, D.V.~Nanopoulos and K.A.~Olive,
{\it No-Scale Supergravity Realization of the Starobinsky Model of
Inflation,} {\sl Phys.\ Rev.\ Lett.\  } {\bf 111}, 111301 (2013);
{\sl Erratum-ibid.\ } {\bf 111}, no. 12, 129902 (2013)
[\arxiv{1305.1247}].

\bibitem{eno7} J.~Ellis, D.~Nanopoulos and
K.~Olive, {\it Starobinsky-like Inflationary Models as Avatars of
No-Scale Supergravity}, \jcap{10}{2013}{009} [\arxiv{1307.3537}].
  %%CITATION = ARXIV:1307.3537;%%

\bibitem{alinde}  R.~Kallosh, A.~Linde, and D.~Roest,
{\it Superconformal Inflationary $a$-Attractors,}
\jhep{11}{2013}{198} [\arxiv{1311.0472}].

\bibitem{class} J.~Ellis, D.V.~Nanopoulos, K.A.~Olive and S.~Verner,
{\it A general classification of Starobinsky-like inflationary
avatars of $SU(2,1)/SU(2)\times U(1)$ no-scale supergravity,}
\jhep{03}{2019}{099} [\arxiv{1812.02192}].

\bibitem{tkref} J.J.M.~Carrasco, R.~Kallosh, A.~Linde and D.~Roest,
{\it Hyperbolic geometry of cosmological attractors,} {\sl Phys.\
Rev.\ D }{\bf 92}, no. 4, 041301 (2015) [\arxiv{1504.05557}];
J.J.M.~Carrasco, R.~Kallosh and A.~Linde, {\it $\alpha
$-Attractors: Planck, LHC and Dark Energy,} \jhep{10}{2015}{147}
[\arxiv{1506.01708}].

\bibitem{tkin} J.J.M. Carrasco, R. Kallosh and A. Linde, {\it Cosmological
Attractors and Initial Conditions for Inflation,} {\sl Phys. Rev.
D }{\bf 92}, no.6, 063519 (2015) [\arxiv{1506.00936}].


\bibitem{nsreview} J.~Ellis, M.A.G.~Garcia, N.~Nagata, D.V. Nanopoulos, K.A.~Olive
and S.~Verner, {\it Building Models of Inflation in No-Scale
Supergravity,} {\sl Int. J. Mod. Phys. D} {\bf 29},  16, 2030011
(2020) [\arxiv{2009.01709}].

\bibitem{ellis21} J.~Ellis, M.A.~G.~Garcia, D.V.~Nanopoulos, K.A.~Olive and
S.~Verner, {\it BICEP/Keck constraints on attractor models of
inflation and reheating,} {\sl Phys. Rev. D} \textbf{105}, no.4,
043504 (2022) [\arxiv{2112.04466}].

\bibitem{linde21} R.~Kallosh and A.~Linde,
{\it BICEP/Keck and cosmological attractors,}
\jcapn{12}{2021}{008}{12} [\arxiv{2110.10902}].



\bibitem{plin} Y.~Akrami {\it et al.} [\plk\ Collaboration],
{\it Planck 2018 results. X. Constraints on inflation}, {\sl
Astron. Astrophys. }\textbf{641}, A10 (2020) [\arxiv{1807.06211}].







\bibitem{tmodel} R. Kallosh and A. Linde, {\it Universality Class in Conformal
Inflation,} \jcap{07}{2013}{002} [\arxiv{1306.5220}].




\bibitem{sky} R.~Kallosh and A.~Linde, {\it Escher in the Sky,} {\sl Comptes Rendus Physique }{\bf 16}, 914 (2015)
[\arxiv{1503.06785}].
%  doi:10.1016/j.crhy.2015.07.004

\bibitem{other}
R.~Kallosh and A.~Linde, {\it Superconformal generalizations of
the Starobinsky model,} \jcap{06}{2013}{028} [\arxiv{1306.3214}];
S.V.~Ketov, {\it Starobinsky Model in $N=2$ Supergravity,} {\sl
Phys. Rev. D }\textbf{89}, no.~8, 085042 (2014)
[\arxiv{1402.0626}]; A.~Linde, {\it Single-field
$\alpha$-attractors,} \jcap{05}{2015}{003} [\arxiv{1504.00663}];
Y.~Aldabergenov, A.~Chatrabhuti and H.~Isono, {\it
$\alpha$-attractors from supersymmetry breaking,} {\sl Eur. Phys.
J. C }\textbf{81}, no.~2, 166 (2021) [\arxiv{2009.02203}].


\bibitem{rube} R.~Kallosh, A.~Linde and T.~Rube, {\it General inflaton potentials in supergravity,}
\prd{83}{2011}{043507} [\arxiv{1011.5945}].

\bibitem{su11} C.~Pallis and N.~Toumbas, \textit{Starobinsky-Type Inflation With Products of K\"ahler
Manifolds}, \jcap{05}{2016}{no. 05, 015} [\arxiv{1512.05657}];
C.~Pallis and N.~Toumbas, {\it Starobinsky Inflation: From
Non-SUSY To SUGRA Realizations,} {\sl Adv. High Energy Phys.
}\textbf{2017}, 6759267 (2017) [\arxiv{1612.09202}].
%%CITATION = doi:10.1155/2017/6759267;%%
%%CITATION = doi:10.1088/1475-7516/2016/05/015;%%


\bibitem{R2} A.A.~Starobinsky, {\it  A New Type of Isotropic Cosmological Models Without Singularity,}
{\sl Phys.\ Lett.\ B }{\bf 91}, 99 (1980).


\bibitem{haber} H.E. Haber, {\it  What is the group of conjugate symplectic
matrices?} {\tt\ssz
http://scipp.ucsc.edu/$\sim$haber/index.html\#sec6}

\bibitem{jhep} G.~Lazarides and C.~Pallis, \textit{Shift Symmetry and Higgs Inflation in Supergravity with
Observable Gravitational Waves}, {\sl J. High Energy Phys.} {\bf
11}, 114 (2015) [\arxiv{1508.06682}].
  %%CITATION = doi:10.1007/JHEP11(2015)114;%%


\bibitem{plcp} N.~Aghanim {\it et al.} [\plk\ Collaboration],
{\it Planck 2018 results. VI. Cosmological parameters }{\sl
Astron. Astrophys. }\textbf{641}, A6 (2020) [\arxiv{1807.06209}].


\bibitem{turner} M. S. Turner, {\it Coherent Scalar-Field Oscillations in an
Expanding Universe} {\sl Phys. Rev. D} {\bf 28}, 1243 (1983).

\bibitem{wreh} J.~Martin and C.~Ringeval, {\it First CMB Constraints on the
Inflationary Reheating Temperature,} {\sl Phys. Rev. D}
\textbf{82}, 023511 (2010) [\arxiv{1004.5525}]; J.~Ellis,
M.A.G.~Garcia, D.V.~Nanopoulos and K.A.~Olive, {\it Calculations
of Inflaton Decays and Reheating: with Applications to No-Scale
Inflation Models}, \jcap{07}{2015}{050} [\arxiv{1505.06986}].


\bibitem{univ} C.~Pallis, \textit{Gravitational Waves, $\mu$ Term
\& Leptogenesis from $B-L$ Higgs Inflation in Supergravity}, {\sl
Universe }\textbf{4}, no.1, 13 (2018) [\arxiv{1710.05759}];
C.~Pallis, {\it Unitarity-Safe Models of Non-Minimal Inflation in
Supergravity,} {\sl Eur. Phys. J. C } \textbf{78}, no.12, 1014
(2018) [\arxiv{1807.01154}]; C.~Pallis and Q.~Shafi, {\it
Induced-Gravity GUT-Scale Higgs Inflation in Supergravity,} {\sl
Eur. Phys. J. C }\textbf{78}, no.6, 523 (2018)
[\arxiv{1803.00349}].
%CITATION = doi:10.1140/epjc/s10052-018-5980-0;%%
%CITATION = doi:10.1140/epjc/s10052-018-5980-0;%%
%CITATION = doi:10.3390/universe4010013;%%


\bibitem{pl20} Y.~Akrami \textit{et al.} [{\it Planck} Collaboration],
{\it $Planck$ intermediate results. LVII. Joint Planck LFI and HFI
data processing,} {\sl Astron. Astrophys. }\textbf{643}, A42
(2020) [\arxiv{2007.04997}].


\bibitem{gws1} P.A.R.~Ade \textit{et al.} [{\sc BICEP} and {\it Keck} Collaboration],
{\it Improved Constraints on Primordial Gravitational Waves using
Planck, WMAP, and BICEP/Keck Observations through the 2018
Observing Season,} {\sl Phys. Rev. Lett. }\textbf{127}, no. 15,
151301 (2021) [\arxiv{2110.00483}].



\bibitem{gws} M.~Tristram \textit{et al.}, {\it Improved limits on the tensor-to-scalar
ratio using BICEP and Planck,} {\sl Phys. Rev. Lett. } {\bf 127},
151301 (2021) [\arxiv{2112.07961}].


\bibitem{rocky} E.D. Stewart and D.H. Lyth, {\it A More accurate analytic calculation of the spectrum of
cosmological perturbations produced during inflation}, {\sl Phys.
Lett. B} {\bf 302}, 171 (1993) [{\tt gr-qc/9302019}]; E.W. Kolb
and S.L. Vadas, {\it Relating spectral indices to tensor and
scalar amplitudes in inflation}, {\sl Phys. Rev. D} {\bf 50}, 2479
(1994) [\astroph{9403001}].


\bibitem{bird} E.~Allys \textit{et al.} [{\sc LiteBIRD} collaboration], {\it Probing Cosmic Inflation with the LiteBIRD Cosmic Microwave
Background Polarization Survey}, \arxiv{2202.02773}.

\end{thebibliography}
\end{document}